\documentclass[12pt]{article}
\usepackage{graphicx}
\usepackage{enumerate}
\usepackage{amssymb,amsmath,amsfonts,palatino,amsthm}
\usepackage{amssymb}
\usepackage{epstopdf}

\newcommand{\beq}{\begin{equation}}
\newcommand{\eeq}{\end{equation}}
\newcommand{\bea}{\begin{eqnarray}}
\newcommand{\eea}{\end{eqnarray}}

\newcommand{\half}{\frac{1}{2}}

\newcommand{\Mscale}{{\cal M}_{\mathrm f}}
\newcommand{\sgn}{\mathrm{sgn}}

\begin{document}
\pagestyle{empty}
\begin{center}
\vskip 2cm
{\bf \LARGE  Aspects of noncommutative (1+1)-dimensional black holes}
\vskip 1cc
Jonas R. Mureika \\
{\small \it Department of Physics, Loyola Marymount University, Los Angeles, CA  90045-2659} \\
\small{Email: jmureika@lmu.edu}
\vskip0.5cm

Piero Nicolini \\
{\small \it Frankfurt Institute for Advandced Studies (FIAS), 
Goethe-Universit\"{a}t, 
  Frankfurt am Main, Germany\\
\small \it Institut f\"{u}r Theoretische Physik, 
 Goethe-Universit\"{a}t, 
Frankfurt am Main, Germany} \\

{\small Email: nicolini@th.physik.uni-frankfurt.de
}
\end{center}

\date{\today}
 \begin{abstract}
 We present a comprehensive analysis of the spacetime structure and thermodynamics of $(1+1)-$dimensional black holes in a noncommutative framework.  It is shown that a wider variety of solutions are possible than the commutative case considered previously in the literature.  As expected, the introduction of a minimal length $\sqrt{\theta}$ cures singularity pathologies that plague the standard two-dimensional general relativistic case, where the latter solution is recovered at large length scales.  Depending on the choice of input parameters (black hole mass $M$, cosmological constant $\Lambda$, {\it etc...}), black hole solutions with zero, up to six, horizons are possible.  The associated thermodynamics allows for the either complete evaporation, or the production of black hole remnants.
 
  \noindent 
 {\small \noindent \\ \scriptsize PACS:  04.70.-s, 04.50.Kd, 04.70.Dy}

 \end{abstract}
\newpage
\pagestyle{plain}

\tableofcontents


\section{Introduction}

There has been a renewal of interest in lower-dimensional gravity models, motivated by the idea of dimensional reduction at high energies.  Causal dynamical triangulation (CDT) approaches suggest that, at the shortest distance scales, spectral physics reduces to a fractal structure of dimension $d=2$ \cite{cdt1,cdt2,cdt3}.  This would indicate a possible renormalizable character of gravity in its quantum regime, as recently shown by implementing an effective minimal length in a spacetime manifold \cite{lmpn1,spectral1,spectral2}. Such a paradigm leads to a background-independent framework for emergent spacetime \cite{emergent}.  

Alternatively, a very recent proposal conjectures a framework whereby spatial dimensions vanish at fixed transition energies $\Lambda_i$, at which the universe reduces from $d=i$ dimensions to $d=i-1$ \cite{dejan,rinaldi}.  If the scales $\Lambda_i \geq 1~$TeV, it may be possible to detect such radical behavior in parton collisions at the LHC, where the jet structure will adopt a planar characteristic.  Quantum black hole production in the reduced $(1+1)$-spacetime may also be detectable through precision measurement of branching ratios \cite{landsberg}.  A particularly novel test of such vanishing dimensions relies on the detection of a gravitational wave ``horizon'' at the millihertz scale \cite{jrmds}.  Scales above this threshold reflect the universe in its $(2+1)-$dimensional phase, in which gravity does not have the available degrees of freedom to produce waves.

We thus explore the consequences of introducing a noncommutative structure to a two-dimensional spacetime.

\section{Noncommutative Geometry}
\label{ncg}

Noncommutative geometry (NCG) is the underlying theory which has inspired a lot of recent work  at the frontier between black hole physics and quantum gravity \cite{Nicolini:2008aj}.  On the formal side NCG implies a modification of the conventional algebra of quantum operators, assuming that coordinate operators might fail to commute. The simplest noncommutative relation one can postulate is
\beq
[X^\mu,X^\nu] = i \theta^{\mu \nu} 
\label{noncomm}
\eeq
where $X^\mu$ is the position operator, $\theta^{\mu \nu}$ is a constant antisymmetric tensor with determinant $\theta\equiv|\theta^{\mu \nu}| $,   having units of area.   However the physical premise of NCG rests in the existence of a minimal length, beyond which coordinate resolution is ambiguous. From the above relation (\ref{noncomm}), one may conclude $\sqrt{\theta}$ is the minimal length. Even if the value of the minimal length $\sqrt{\theta}$ is not set by (\ref{noncomm}), on physical grounds we can think $\sqrt{\theta}$ to be the Planck length or very close to it, i.e. $\sqrt{\theta}\sim 10^{-33}$ cm. The tininess of this scale would imply that NCG can be probed only in extreme energy phenomena. On the other hand, in the present of extradimensions, the new fundamental scale $\Mscale$ can be closer to current particle physics experiments: in the light of Cavendish experiment data about possible deviations of Newton's law we can set $1/\sqrt{\theta}\sim \Mscale\sim 1$ TeV \cite{newton,newton2}.

In spite of this consistent mathematical assumption and the physically sound motivation, NCG is  hard to implement in the context of gravity. Indeed the formulation of noncommutative gravity, the theory of gravitation in terms of noncommutative coordinates, is far from being complete. Noncommutative gravity is dual to a theory in which the conventional product among vielbein fields is replaced by a nonlocal Weyl-Wigner-Moyal $\star$-product  \cite{Moffat:2000gr,Calmet:2005qm}. Modifications  to the Einstein-Hilbert action due to the $\star$-product are obtained by expanding the nonlocal product in powers of $\theta$ \cite{Calmet:2006iz}. However at any truncation of the expansion at a given order, one finds corrections which are unable to cure the conventional bad short distance behavior of black hole spacetimes \cite{Chai,Chai2,saha,kob}.  Against this background, there exists the possibility of adopting an effective approach, by focusing on the specific feature of noncommutative smearing: in the absence of a resolution beyond the minimal length $\sqrt{\theta}$, a pointlike object is no longer meaningful in a noncommutative background. It has been shown that the sharpest distribution which faithfully described a localized object is no longer a Dirac delta, but a Gaussian, whose width coincides with $\sqrt{\theta}$ 
\begin{equation}
f(\vec{x}) = \frac{1}{(4\pi \theta)^{n/2}} \exp\left({\frac{-{\vec{x}}^2}{4\theta}}\right)
\label{gp}
\end{equation}
where $n$ is the manifold dimension \cite{Smailagic:2003yb,Smailagic:2003rp,Smailagic:2004yy,Spallucci:2006zj,Banerjee:2009xx}.
In addition one can show that primary corrections to gravity field equations in the presence of a noncommutative smearing can be obtained by replacing a pointlike source term (matter sector) with a Gaussian distribution, while keeping formally unchanged differential operators (geometry sector) \cite{Nicolini:2005zi}. More specifically this is equivalent to saying that the only modification occurs at the level of the energy-momentum tensor,  while the Einstein tensor is formally left unchanged. Recently it has been shown that Einstein gravity coupled to an effective energy momentum tensor of this kind is dual to a ultraviolet complete quantum gravity in which a nonlocal Einstein-Hilbert action is coupled to ordinary matter source terms \cite{Modesto:2010uh}. As a consequence the Gaussian profile $f(\vec{x})$ is the result of the action of a nonlocal operator $e^{\theta\Box}$ on the Dirac delta function,
\begin{equation}
f(\vec{x})= e^{\theta \Box}\delta(\vec{x})~~.
\label{gp}
\end{equation}
For the static spherically symmetric case one has to solve Einstein equations  with ``smeared'' energy density profile $\rho_\theta(\vec{x})=Mf(\vec{x})$ and non-vanishing pressure terms, i.e.,
\beq
T^r_r = -T^0_0 = -\rho_\theta(r)~~~,~~~p_{\perp} = -\rho_\theta(r) -\frac{r}{2} \partial_r \rho_\theta(r)
\eeq
which describes a self-gravitating anisotropic fluid.  The condition $T^r_r = -T^0_0$ guarantees the usual Schwarzschild symmetry in the metric coefficients $g_{00} = -g_{rr}^{-1}$, while tangential pressures are determined via the divergence-free condition $\nabla_\nu T^{\mu \nu} = 0$.
As a result one finds the noncommutative geometry inspired Schwarzschild line element \cite{nss}
\begin{equation}
ds^2=-\left(1-\frac{2MG}{r}\ \frac{\gamma(3/2;r^2/4\theta)}{\Gamma(3/2) }\right)dt^2+\frac{dr^2}{\left(1-\frac{2MG}{r}\ \frac{\gamma(3/2;r^2/4\theta)}{\Gamma(3/2) }\right)}+r^2d\Omega^2
\label{ncschw}
\end{equation}
where $\Gamma(3/2)=\sqrt{\pi}/2$ and 
\begin{equation}
\gamma(3/2;r^2/4\theta)=\int_0^{r^2/4\theta}dt \ t^{1/2}e^{-t}.
\end{equation}
We note that for $r\gg \sqrt{\theta}$ the ratio $\gamma/\Gamma\to 1$ and (\ref{ncschw}) reproduces the Schwarzschild geometry. On the other hand for $r\ll \sqrt{\theta}$ the ratio $\gamma/\Gamma\sim r^3$ and (\ref{ncschw}) turns out to be an everywhere regular geometry with a deSitter core in place of the curvature singularity at the origin.
Additional black hole solutions have been examined for NCG, including charged \cite{charged}, rotating \cite{rotating}, charged-rotating  \cite{crotating}, dirty \cite{dirty} and Schwarzschild-deSitter \cite{SchwdS} cases. In the presence of extradimensions further solutions have been determined both for the neutral \cite{higher},  charged \cite{chargedhigher}, rotating  and charged-rotating  cases \cite{crotating}.
All these improved spacetimes show that NCG can incorporate model independent characters which are expected in several quantum gravity modifications of spacetime geometries \cite{lm1,lm2,lm3,lm4,lm5,lm6,br1,br2,br3}.

\section{A (1+1)-Dimensional Black Hole Framework}
\label{11bh}
Prior research in lower-dimensional gravity has led to breakthroughs in quantum gravity (for a literature review, the reader is directed to \cite{collas,henneaux,brown,mann1,mann2,mannsik,mann3,mann4,balbinot1,balbinot2,kummer}).    Since the Ricci tensor is a topological invariant in two-dimensional spacetime, the corresponding action must be different than Einstein-Hilbert.  It is generally most common to introduce to the action a dilaton field, which yields a set of field equations  coupled to the field's evolution.

Perhaps the simplest extrapolation of Einstein gravity in (1+1)-dimensions can be extracted from the general action \cite{mann3,mann4} 
\beq
S[g_{\mu \nu}, \phi] = \int d^2x~\sqrt{-g} \left[ \frac{1}{16\pi G }\left(\psi R+\Lambda+\frac{1}{2}(\nabla \psi)^2\right) +{\cal L}_m\right]
\label{2daction}
\eeq
where $R$ is the Ricci, $\psi$ is a scalar dilatonic field and ${\cal L}_m$ is the matter Lagrangian.
The variation with respect to $\psi$ gives
\bea 
 R - \nabla^2\psi = 0  \label{eq1}
\eea
while that with respect to $g_{\mu\nu}$ gives
\bea
\frac{1}{2}\nabla_\mu \Psi \nabla_\nu \Psi - g_{\mu \nu} \left(\frac{1}{4} \nabla^\lambda \Psi \nabla _\lambda \Psi - \nabla^2 \psi\right) -\nabla_\mu \nabla_\nu \Psi = 8\pi G T_{\mu \nu} + \frac{1}{2}\Lambda g_{\mu \nu} 
 \label{eeq11} 
\eea
where the stress-energy tensor is 
\beq 
T^{\mu\nu} = -\frac{2}{\sqrt{-g}}\frac{\delta {\cal L}_m}{\delta g_{\mu\nu}}.
\eeq
By tracing (\ref{eeq11}), the dilaton $\psi$ decouples from Einstein equations, simplifying (\ref{eq1}) to the standard Liouville gravity form
\begin{equation}
R-\Lambda=8\pi G T
\label{eq2}
\end{equation}
where $T=T^\mu _\mu$. 

By prescribing the energy momentum tensor $T^{\mu\nu}$ one gets a variety of solutions with nontrivial horizon structure. For instance, in the case of a static source representing a point particle at the origin the energy density\footnote{The general representation of the Dirac delta in a arbitrary coordinate system $\left\{q^i\right\}$ requires the inclusion of a term depending on the Jacobian, i.e., $\delta(q^1, q^2, \dots)=\frac{\delta(q^1)\delta(q^2)\dots}{{\cal J}}$, where ${\cal J}=\left|\frac{\partial {\mathrm x}^i}{\partial q^j}\right|$ and ${\mathrm x}^i$ are Cartesian coordinates. The authors in \cite{mann2} ignored the Jacobian in describing the point like static source to simplify Einstein equations. This is based on the fact that ${\cal J}$ is everywhere finite and asymptotically approaches the unit. As a result the absence of ${\cal J}$ does not modify the main features of the solution at least at the leading order.} 
\beq
\rho = \frac{M}{2\pi G}\delta(x)
\label{pointmass}
\eeq
where the parameter $M$ corresponds to the ADM mass, while the pressure is $p =0$. 
For the generic line element
\begin{equation}
ds^2=-\alpha(x)\ dt^2+\alpha^{-1}(x)\ dx^2
\end{equation}
the equations of motion  (\ref{eq2}) read
\beq
\frac{d^2}{dx^2} \alpha(x) + \Lambda = 4M\delta(x).
\label{2deeq}
\eeq
Assuming a linearly-symmetric solution about the origin ($\alpha = \alpha(|x|)$), this becomes 
\beq
\alpha^{\prime \prime} + 2\alpha^\prime \; \delta(x)+ \Lambda = 4M\delta(x)~~,
\eeq
where prime denotes $d/d|x|$. If $\alpha$ is continuous then one is led to the consistency condition $2\alpha^\prime = 4M$ at $x=0$.
It can be shown \cite{mann3} that the solution is $\alpha(x) = \frac{1}{2} \Lambda x^2+2M|x| -C$, and so the two-dimensional metric is
\beq
ds^2 = -\left(- \half \Lambda x^2 + 2 M |x| -C\right) dt^2 + \frac{dx^2}{\left(-\half  \Lambda x^2 + 2 M|x| -C\right)}.
\label{11metric}
\eeq
The constants $M$, $\Lambda$ and $C$  are arbitrary, and their values determine several distinct classes of solutions and relative causal structures \cite{manndan}.  The sign convention for $\Lambda=\pm|\Lambda|$  is deSitter ($-$) and anti-deSitter ($+$). For $M>0$ the above metric describes the 2-dimensional analogue of a Schwarzschild black hole  \cite{mann2}.
If $\Lambda\neq 0$ we have an outer $|x|_>$ and a inner $|x|_<$ horizon at 
\begin{equation}
|x|_{\lessgtr}=\frac{2M\pm\sqrt{4M^2- 2\Lambda C}}{\Lambda},
\end{equation}
only if $2M^2>C\Lambda$. For $2M^2=C\Lambda$, we have a single degenerate horizon at $|x|_0= 2M/\Lambda$, which exists only if $M$ and $\Lambda$ have the same sign.
For $\Lambda= 0$ we have just one horizon at 
\begin{equation}
|x|_1=\frac{C}{2M}
\end{equation}
which exists only if $M$ and $C$ have the same sign.

  \begin{figure}
 \begin{center}
 \includegraphics[height=5.8cm]{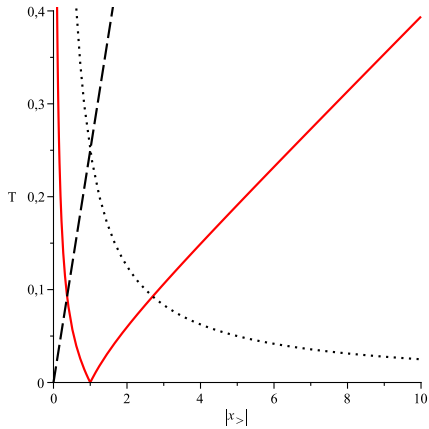}
 
     \caption{\label{figT}  The black hole temperature as a function of the horizon radius $|x_>|$ meant as a dimensionless variable. The solid curve is the two horizon case $2\pi T/ \hbar=\left|-\frac{1}{4} |x|_>+\frac{1}{4|x|_>}\right|
$ for  $C=1/2$ and $\Lambda=1$, which describes the temperature for $|x_>|\geq 1$ since $|x_0|=1$ in our units. The dotted curve corresponds to the single horizon case $2\pi T/ \hbar=\frac{1}{4|x|_>}$, namely for $C=1/2$ and $\Lambda=0$. Finally the dashed curve, $2\pi T/ \hbar=\left|-\frac{1}{4} |x|_>\right|
$ and $C=0$ and $\Lambda=1$,  is a special case with two horizons, one of these at the origin.}
     \end{center}
  \end{figure}

The Hawking temperature of the two-dimensional black hole solution can be calculated via the usual Wick rotation (\ref{11metric}),
\beq
ds^2 = \alpha(x)~d\tau^2 + \alpha^{-1}(x)~dx^2 ~~\longrightarrow ~~ \alpha(x(r))~d\tau^2 + dr^2~~,\alpha(x) = \left(\frac{dx}{dr}\right)^2
\eeq
The periodicity of $\alpha$ yields the standard temperature at the horizon $x_H$
\beq
T = \frac{\hbar}{2\pi} \left|\frac{\alpha^\prime(x_H)}{2}\right| = \frac{\hbar}{2\pi}\sqrt{M^2 - C\frac{\Lambda}{2}}
\eeq
where the latter equality follows upon using (\ref{11metric}).  Note that unlike the standard (3+1)- and higher dimensional cases, the temperature varies linearly with mass \cite{mann2}. We also note that the degenerate case with a single horizon, i.e., $|x|_0= 2M/\Lambda$ actually corresponds to a zero temperature extremal configuration. We should better say that the latter is the equilibrium configuration between the thermalizing effect of the black hole and the background cosmological bath. In this scenario the hot horizon $|x|_>$ by emitting radiation, shrinks, reaches the final equilibrium where its evaporation stops, i.e., $|x|_0$ and leaves a stable remnant with mass $M_0=\pm \sqrt{|\Lambda C|/2}$. On the contrary for $\Lambda=0$, we have simply
\beq
T = \frac{\hbar}{2\pi} \left|M \right|,
\eeq
and the evaporation ends up by exhausting the mass $M$.
For later convenience we write the temperature as a function of $|x|_>$ as
\begin{equation}
T=\frac{\hbar}{2\pi}\left|-\frac{1}{4}\Lambda |x|_>+\frac{C}{2|x|_>}\right|
\label{temp}
\end{equation}
which is defined for $|x|_>\geq|x|_0$ and vanishes for $|x|_> = |x|_0=\sqrt{2|C/\Lambda|}$.

An alternate class of two-dimensional black hole solutions can be obtained from the CGHS formalism (see {\it e.g.} \cite{cghs1,cghs2,cghs3,cghs4}), whose action is of the form

\beq
S[g,\phi,f] = \frac{1}{G} \int d^2x~\sqrt{|g|} e^{-2\phi} (R + 4 \nabla^a \phi \nabla_a \phi + 4\kappa^2) -\frac{1}{2} \int d^2x ~ \sqrt{|g|} \nabla^a f \nabla_a f
\eeq
which is a functional of the metric $g_{ab}$, a dilaton field $\phi$, and a (scalar) matter field $f$.  The constants $G$ and $\kappa$ are the two-dimensional equivalents of Newton's constant and the cosmological constant.  The decoupling of the matter field from the dilaton and curvature tensor allows aspects of the theory to depend entirely on the behavior of $f$.  Unlike the action considered herein, the CGHS black hole is not a formally-geometric construct, but instead arises from the dynamics of the interacting fields.  In particular, the gravitation collapse behavior of $f$ induces the the associated black hole solutions on the spacetime manifold.  The similarity of the above action to a four-dimensional analog indicates that CGHS black hole solutions can yield valuable information about the classical collapse and quantum evaporation behavior of their higher-dimensional counterparts \cite{cghs3,cghs4}.

\section{A noncommutative black hole in two dimensions}
Previous studies have already addressed non-commutative black holes in (1+1) - dimensions.   In \cite{pn2}, a radiating toy, i.e., (1+1) dimensional Schwarzschild black hole is studied to demonstrate that there exists a maximum temperature profile, as well as a non-zero mass remnant phase.  Similarly, the trace anomaly stemming from coordinate fluctuations has been examined, specifically in the Polyakov action and associated two-dimensional $r-t$ slice of a Schwarzschild-like black hole solution \cite{Spallucci:2006zj}.

Here, we consider the non-commutative limit of the black hole described in \cite{mann2}.  Combining the prescriptions of Sections~\ref{ncg} and \ref{11bh}, we now assume the spacetime is endowed with a minimal length $\sqrt{\theta}$.  We postulate the metric solution to be a variant of the form expressed in Eq.~(\ref{11metric}), and thus will have the form
\beq
ds^2 = -\alpha_\theta(x)\; dt^2 + \frac{dx^2}{\alpha_\theta(x)}~~~.
\eeq
The mass distribution will be  affected by the noncommutative smearing. According to the tenets of noncommutative geometry the leading modifications to the energy density are of the kind
\beq
\rho(x)=\left(\frac{M}{2\pi G}\right) \delta(x) ~~\longrightarrow~~\rho_{\theta}(x) =\left( \frac{M}{2\pi G}\right)\frac{1}{\sqrt{4\pi \theta}} \exp\left(-\frac{x^2}{4\theta}\right)
\label{ncdensity}
\eeq
such that the ADM mass is
\beq
M = 2\pi G \int_{-\infty}^\infty dx\; \left|\frac{\partial \mathrm{x}}{\partial x}\right|\rho_{\theta}(x).
\label{ncmass}
\eeq
where $\mathrm{x}$ is a Cartesian-like freely-falling coordinate. The above mass parameter represents the whole mass-energy available in the spacetime. It has an equivalent meaning of the mass used in the solution \cite{mann2,mann3}. However, as it happens for the noncommutative geometry inspired line element in four dimensions, different values of $M$ determine a variety of properties of the solution.
To justify the prescription (\ref{ncdensity})  we follow the route of the nonlocal deformation of the action (\ref{2daction}) by means of the operator $e^{\theta\Box}$ \cite{Modesto:2010uh,jm10}, i.e., 
\begin{equation}
S[g_{\mu \nu}, \phi] = \int d^2x~\sqrt{-g} \left[ \frac{1}{16\pi G }\left(\psi {\cal R}(x)+\Lambda+\frac{1}{2}(\nabla \psi)^2\right) +{\cal L}_m\right],
\end{equation}
where
\begin{equation}
{\cal R}(x)=\int d^2y~\sqrt{-g} \ {\cal A}(x-y) R(y)
\end{equation}
with 
\begin{equation}
{\cal A}(x-y)=e^{-\theta\Box}\ \delta^{(2)}(x-y).
\end{equation}
By neglecting surface terms coming from the variation of $\Box$, one gets the following equations
\begin{equation}
e^{-\theta\Box} R -\Lambda=8\pi GT,
\label{nleq}
\end{equation}
that can be cast in the more conventional form as 
\begin{equation}
 R -\Lambda =8\pi G e^{\theta\Box}T.
 \label{nceq}
\end{equation}
As expected the nonlocal deformations of the geometry coupled to a classical source term (\ref{nleq}) are equivalent to a classical geometry  coupled to a noncommutative  smeared source term (\ref{nceq}). As a result we are left with the calculation of the new profile for the energy density. 
We temporarily switch to the Cartesian-like free-falling coordinates $(\mathrm{x}, \mathrm{t})$ for computational
convenience and we write
\begin{eqnarray}
\rho_{\theta}(\mathrm{x}) &\equiv & e^{\theta\Box}\rho_{\theta}(\mathrm{x})\\
&= &\left( \frac{M}{2\pi G}\right) \ e^{\theta \partial_{\mathrm x}^2}\ \delta(\mathrm{x}) \\
&=&
\left( \frac{M}{2\pi G}\right)\frac{1}{2\pi} \int_{-\infty}^\infty dp \ e^{-\theta p^2} e^{i\mathrm{x}p} \\
&=&\left( \frac{M}{2\pi G}\right)\frac{1}{\sqrt{4\pi \theta}} \ e^{-\mathrm{x}^2/4\theta}
\end{eqnarray}
which has almost the desired shape. The final step is to transform back to the original coordinates $(\mathrm{x},\mathrm{t})\to(x,t)$. However, if $\alpha_\theta (x)$ is  a regular function at the leading order one still finds
\beq
\rho_{\theta}(x) = \left( \frac{M}{2\pi G}\right)\frac{1}{\sqrt{4\pi \theta}} \exp\left(-\frac{x^2}{4\theta}\right)+\dots
\label{ncdensity2}
\eeq
where dots stand for subleading corrections in curvature tensors. We can conclude that within the approximations in which noncommutative geometry works the Gaussian profile replaces the Dirac delta function.

With this fluid type energy density profile we must consider the correct form for the pressure term in order to have a conserved energy momentum tensor. For a perfect fluid one can write
\begin{equation}
T^{\mu\nu}=(\rho_\theta +p)u^\mu u^\nu +p g^{\mu\nu}.
\end{equation}
In case of a fluid at rest in a time-like region we have $u^0=1/\sqrt{\alpha_\theta(x)}$ and $u^1=0$ and (\ref{eeq11}) reads
\begin{equation}
\frac{d^2}{dx^2}\alpha_\theta(x)+\Lambda =8\pi G (\rho-p).
\label{ncfe}
\end{equation}
The field equation above must be solved in conjunction with the equation of hydrostatic equilibrium
\begin{equation}
\frac{dp}{dx}=-\frac{1}{2}\left(\frac{d}{dx}\ln(\alpha)\right)\left(\rho_\theta+p\right)
\label{cons}
\end{equation}
which follows from the covariant conservation of $T^{\mu\nu}$ when $p=p(x)$ and $\rho_\theta=\rho_\theta(x)$. Since $\rho_\theta$ is specified one can insert (\ref{cons}) into (\ref{nceq}) and get 
\begin{equation}
\ddot{\alpha}(p^\prime)^2+\dot{\alpha}(p^{\prime\prime})+\Lambda =8\pi G(\rho_\theta -p)
\end{equation}
where $\dot{\alpha}\equiv d\alpha/dp$. The above equation can be solved for $p(x)$, which in
turn allows one to solve for $\alpha$. Requiring a symmetry solution about $x=0$, one can formally write
\begin{equation}
\alpha(|x|)=\alpha(0)\exp\left(-2\int_0^{|x|} \left(\rho_\theta(z)+p(z)\right)^{-1}\frac{dp(z)}{dz}\ dz\right).
\end{equation}
The solution of the above equation is hard to get in analytical exact form. 
 However we know that at large distances, i.e., $|x|\gg\sqrt{\theta}$, the energy density is vanishing, $\rho_\theta\approx 0$, while at small scales, i.e., $|x|\ll\sqrt{\theta}$, it is a constant, $\rho_\theta\approx \left( \frac{M}{2\pi G}\right)\frac{1}{\sqrt{4\pi \theta}}$. In addition if we write (\ref{nceq}) in terms of $|x|$ we get
\beq
\alpha^{\prime \prime} + 2\alpha^\prime \; \delta(x)+ \Lambda = 8\pi G (\rho_\theta - p).
\eeq
If $\alpha$ is continuous then one is led to a new consistency condition $\alpha^\prime = 0$ at $x=0$.
Thus (\ref{cons}) is solved if $p(x)\approx 0$. The pressure is nonzero only in an intermediate region between the small and the large scale regime and provides the outward stress which prevents the Gaussian energy profile to collapse in a Dirac delta. Since the effects of noncommutative geometry are relevant at short scales and the region for which $p(x)$ is nonzero is limited, we can neglect it to solve (\ref{cons}) and (\ref{nceq}) at the leading order.

As a consequence, we can integrate the Einstein equations (\ref{2deeq}) to get 
\beq
\alpha_\theta(x) = -\frac{1}{2}\Lambda x^2 +\frac{4 M\sqrt{\theta}}{\sqrt{\pi}}
\left(
 \exp\left(-\frac{x^2}{4\theta}\right)
+\frac{1}{2\sqrt{ \theta}}  \gamma\left(\half,\frac{x^2}{4\theta}\right)
|x|\right)  - C
\label{ncsolution}
\eeq
where 
\begin{equation}
\gamma\left(\half,\frac{x^2}{4\theta}\right)=\int_0^{x^2/4\theta}dt\ t^{-1/2}\ e^{-t}
\end{equation}
is the incomplete gamma function. The behaviour of $\gamma(1/2; x^2/4\theta)$ is crucial. At large distances, $|x|\gg\sqrt{\theta}$, the gamma function is $\gamma\approx \sqrt{\pi}-2\sqrt{\theta}\exp(-x^2/4\theta)/|x|$. This means that, as expected (\ref{ncsolution}) matches the usual $\alpha(x)$. Conversely new physics emerges at short distances. Since for $|x|\ll\sqrt{\theta}$, the gamma function is $\gamma\approx |x|/\sqrt{\theta}$ we get
\begin{equation}
\alpha_\theta(x)\approx \frac{4M\sqrt{\theta}}{\sqrt{\pi}}-C+\left(\frac{M}{\sqrt{\pi\theta}} -\frac{1}{2}\Lambda \right)\  x^2.
\end{equation}
This means that noncommutativity has smoothed the local profile of the solution which has a quadratic dependence on $|x|$, instead of a linear one. As a consequence the Ricci scalar, $R=-\frac{d^2}{dx^2} \alpha(x)$, turns out the be everywhere finite.

 \begin{figure}
 \begin{center}
 \includegraphics[height=4.0cm]{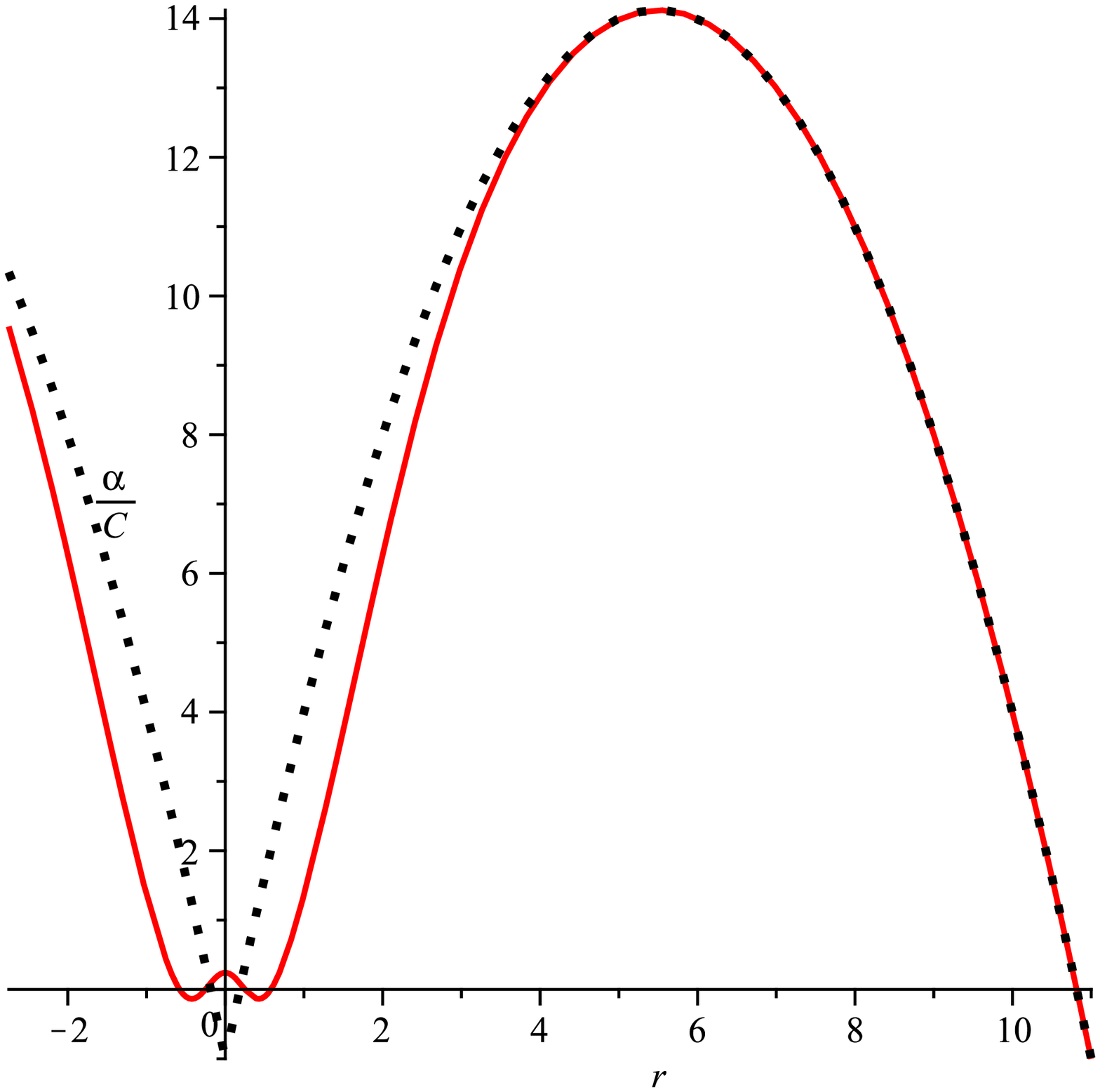}
  \includegraphics[height=4.0cm]{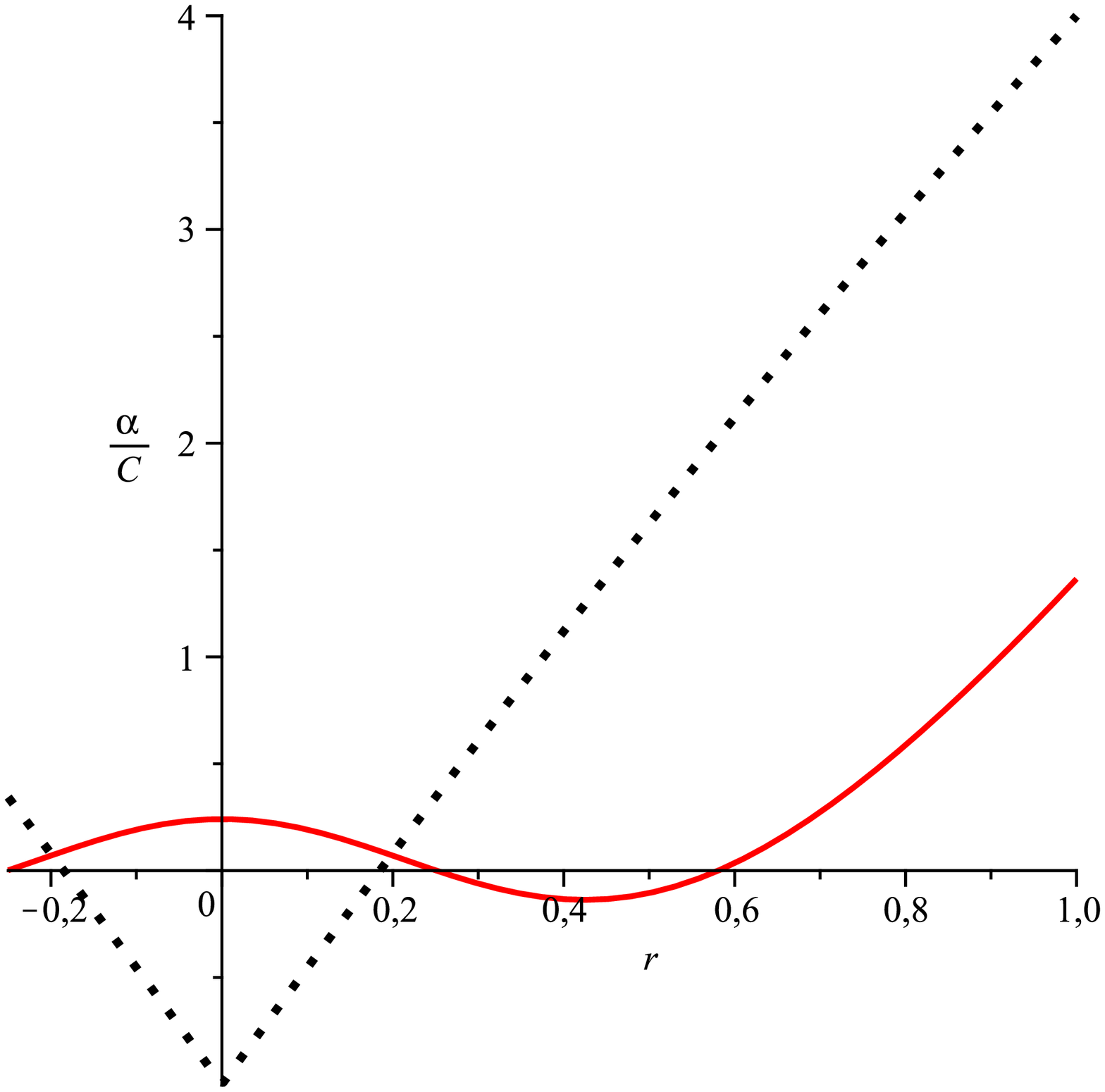}
   \includegraphics[height=4.0cm]{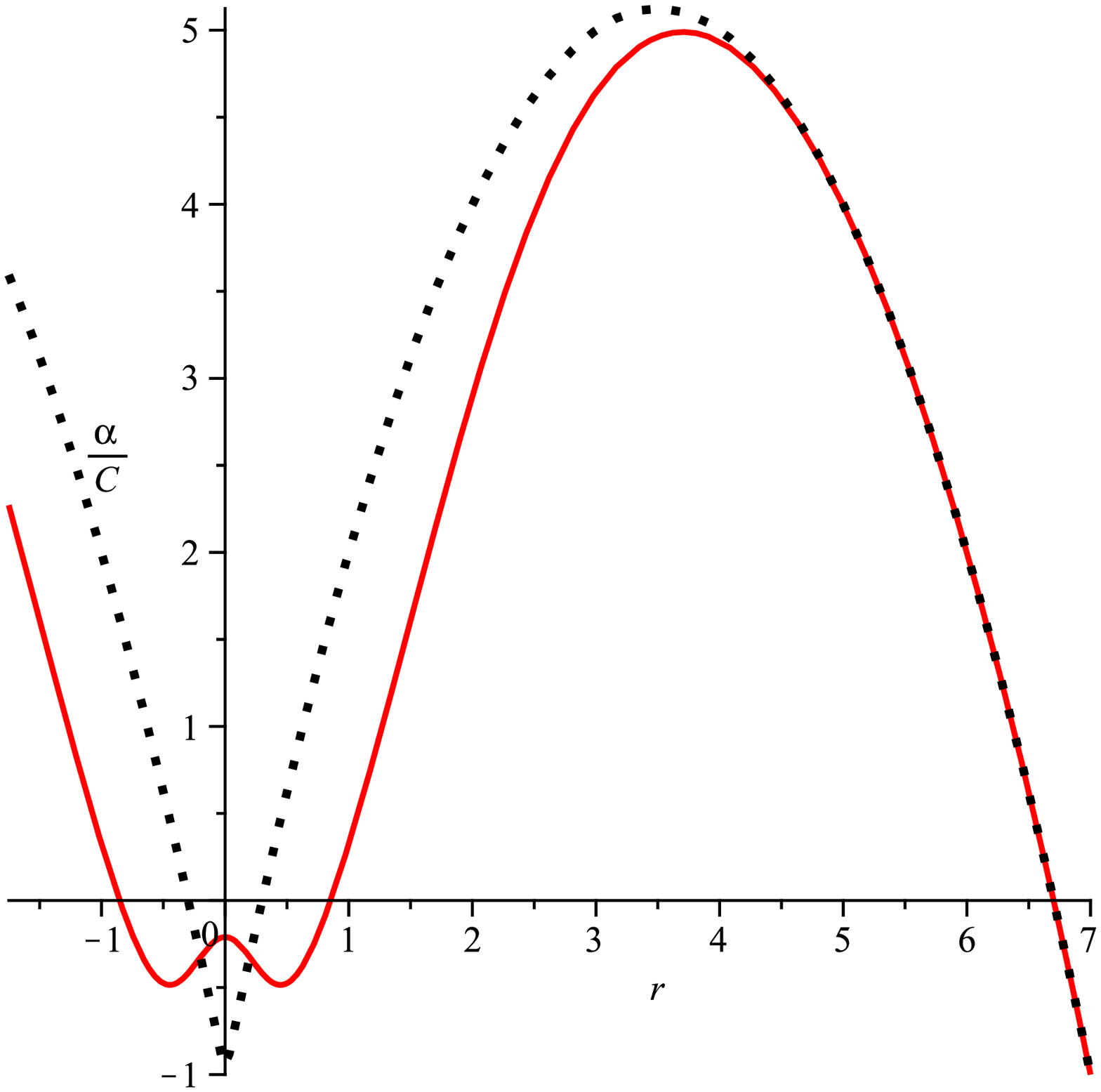}
     \includegraphics[height=4.0cm]{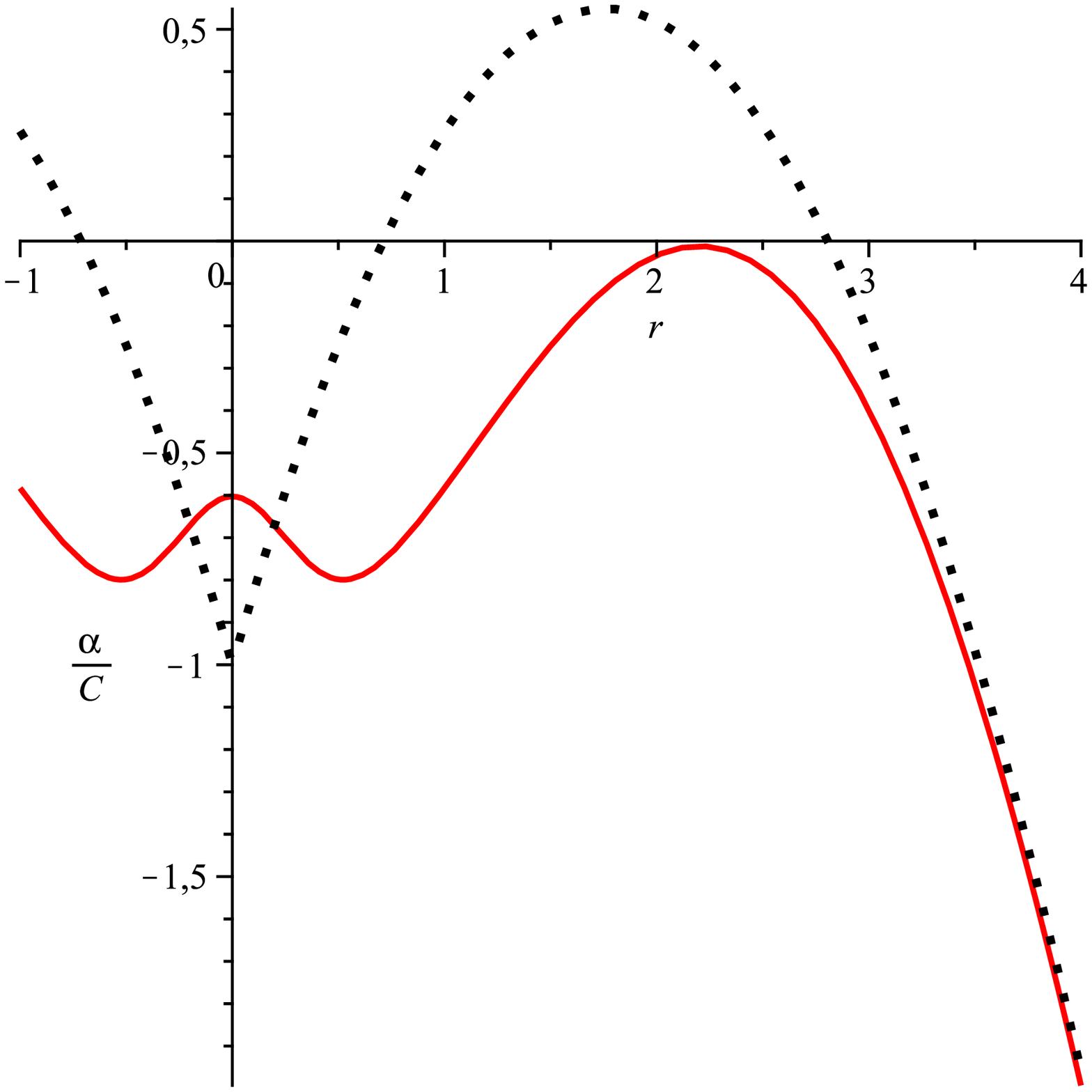}
      \caption{\label{figure1} Behavior of $\alpha_\theta(x)$ for parameter values (from left):
      $q=5$, $m=0.55$; $q=5$, $m=0.55$; $q=5$, $m=0.35$; and $q=5$, $m=0.176$.  The dotted plot is the generic commutative solution for the parameters in question.
       The wide array of causal structures is evident: the left-most figure contains as many as six horizons, while the right-most contains two (degenerate) ones.
      } 
     \end{center}
  \end{figure}  

Concerning the study of the horizon we may wish to solve the equation $\alpha_\theta(x_H)=0$. However we  cannot solve this equation in a closed form. It is worthwhile to write the line element as
\begin{equation}
\frac{\alpha_\theta(r)}{C}=-1\mp\frac{1}{2}\ r^2+\frac{4m}{\sqrt{\pi}} \left[ \exp\left(-\frac{q^2 r^2}{4}\right)
+\frac{1}{2} (q|r|)  \gamma\left(\half,\frac{q^2r^2}{4}\right)\right]
\end{equation}
where $r=\sqrt{|\Lambda/C|}\ x$, $m=\sqrt{ \theta}M/C$ and $q=\sqrt{|C|/\theta|\Lambda|}$. The sign $-$ in front of $\half r^2$ occurs when $\Lambda/C>0$ and conversely we have $+$ for $\Lambda/C<0$. The parameter $\sqrt{\theta}$ sets the unit of length for plots, and $q$ gauges the extent of the region affected by noncommutative effects at length scale associated with the cosmological constant term. Indeed for $q\gg 1$ one gets
\begin{equation}
\frac{\alpha_\theta(r)}{C}\approx -1\mp\frac{1}{2}\ r^2+2m q|r|=-1\pm\frac{1}{2}\left|\frac{\Lambda}{C}\right|x^2 +2\frac{M}{C}|x|
\end{equation}
which is nothing but the line element (\ref{11metric}), with the usual horizon equation.  However the actual scale which controls noncommutative effects is $q|r|$. We can see this by the fact that even for large $q$ if $r$ is small, i.e., $|r|\sim 1/q$, the spacetime has a different behaviour with respect to (\ref{11metric}). In Fig.  \ref{figure1}, we have the spacetime for $\Lambda/C>0$ and $m>0$ which has a smooth behavior at the origin, giving rise to the possibility of an additional inner horizon. For smaller values of $m$ the curve lowers down, giving rise to a single horizon geometry and eventually to a no horizon geometry. 

The rich causal structure is evident from these figures.  The left-most figure admits up to six horizons with an overall external spacelike structure surrounding a time-like region, an innner spacelike region, and a ``core'' timelike one.  The figure on the far right, on the other hand, shows a global spacelike structure with a possible degenerate set of horizons at $|r| \sim 2.2$.

\begin{figure}
 \begin{center}
 \includegraphics[height=4.0cm]{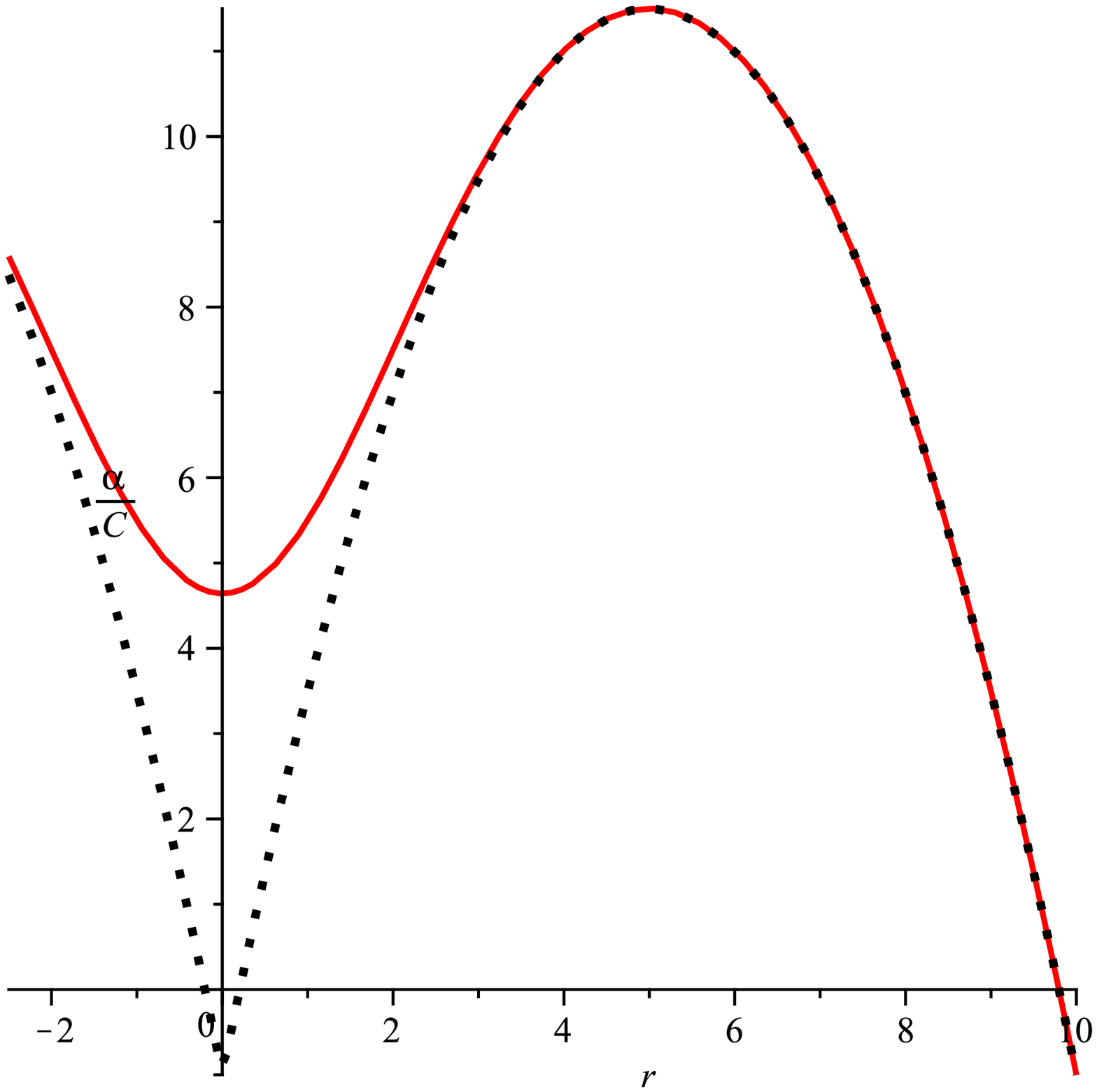}
   \includegraphics[height=4.0cm]{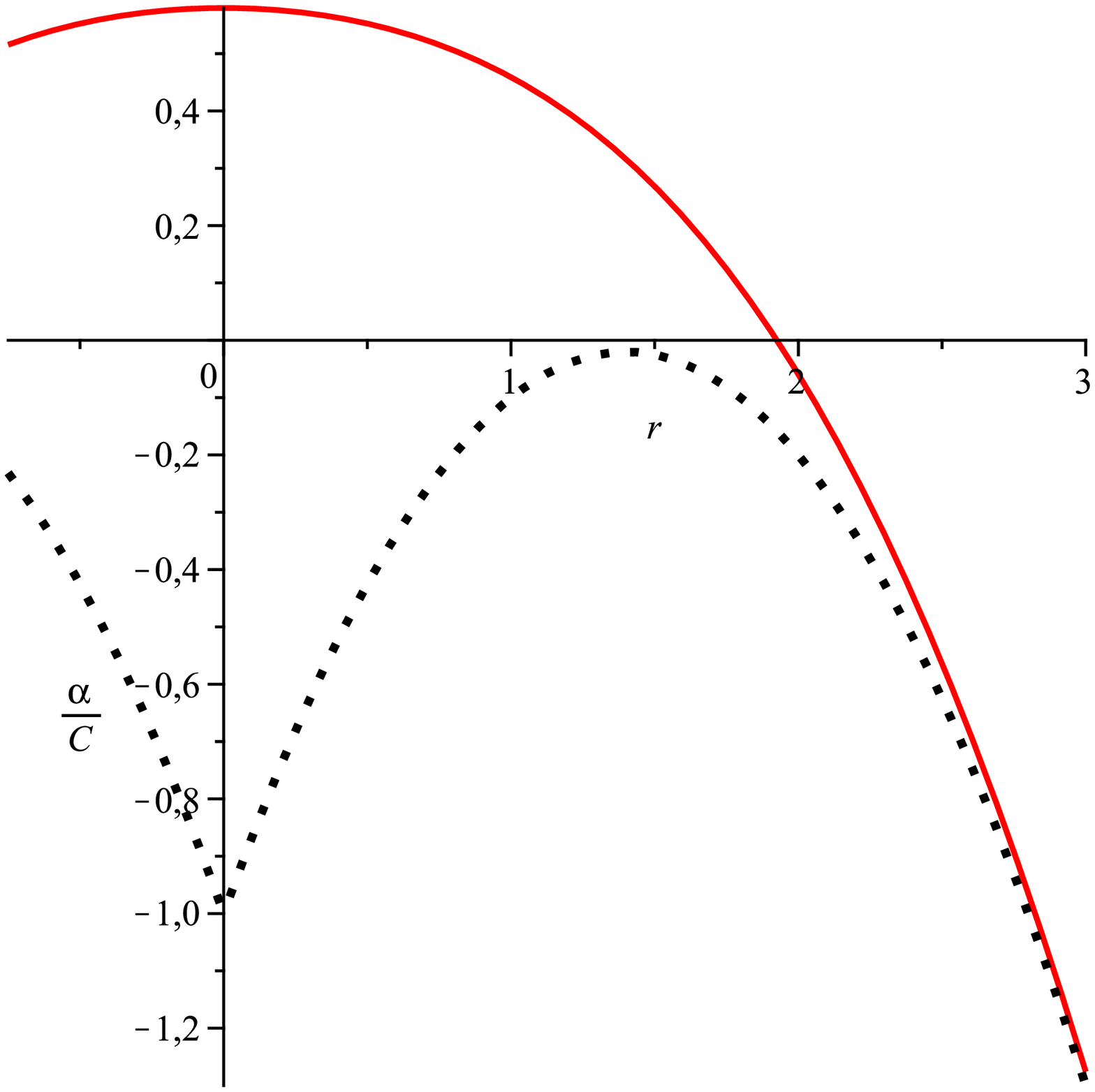}
      \caption{\label{figure2}
Behavior of $\alpha_\theta(x)$ for parameter values (from left):
      $q=1$, $m=2.5$;  and $q=1$, $m=0.7$.  The dotted plot is the generic commutative solution for the parameters in question. }
     \end{center}
  \end{figure}

\begin{figure}
 \begin{center}
 \includegraphics[height=4.0cm]{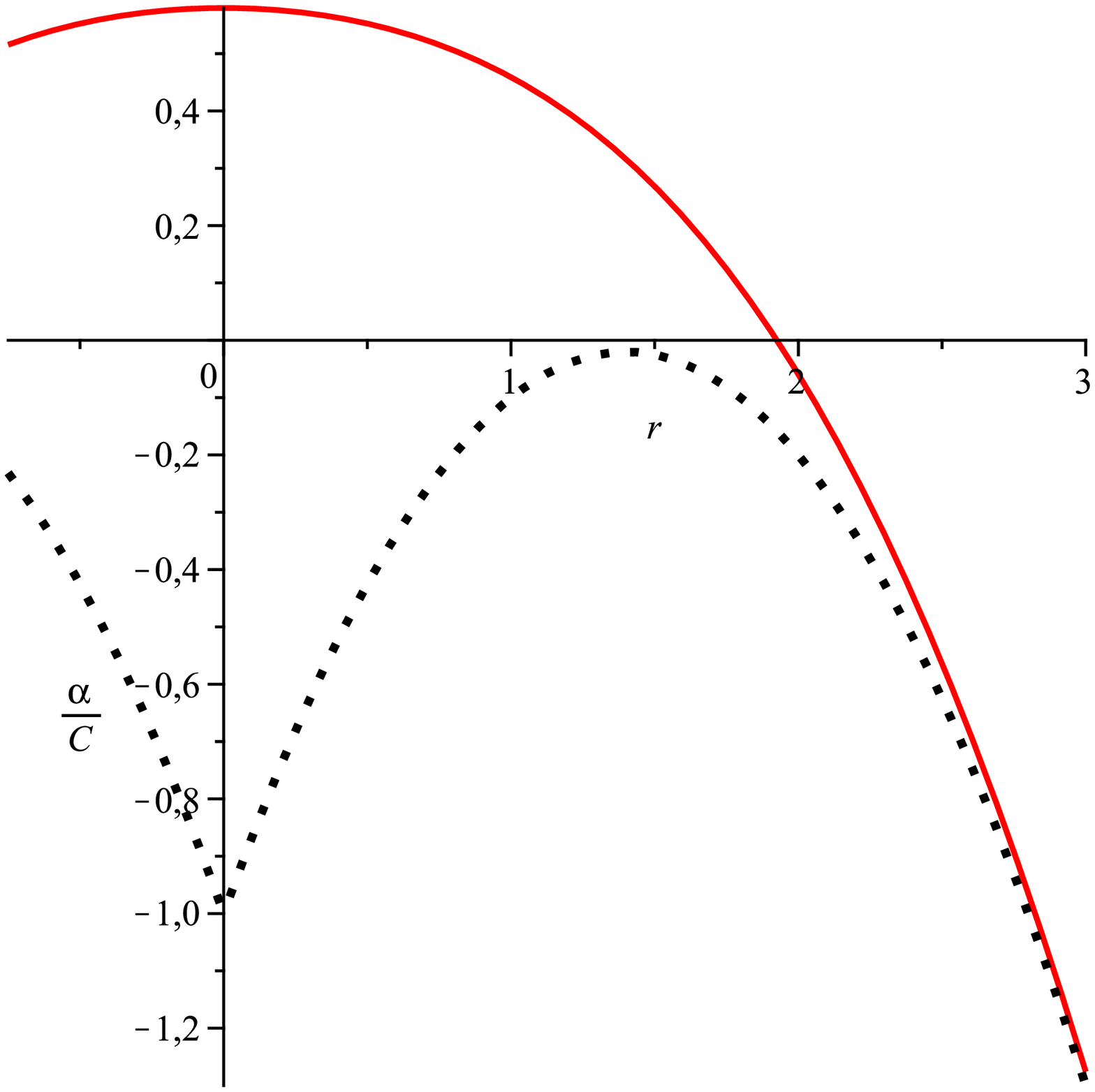}
  \includegraphics[height=4.0cm]{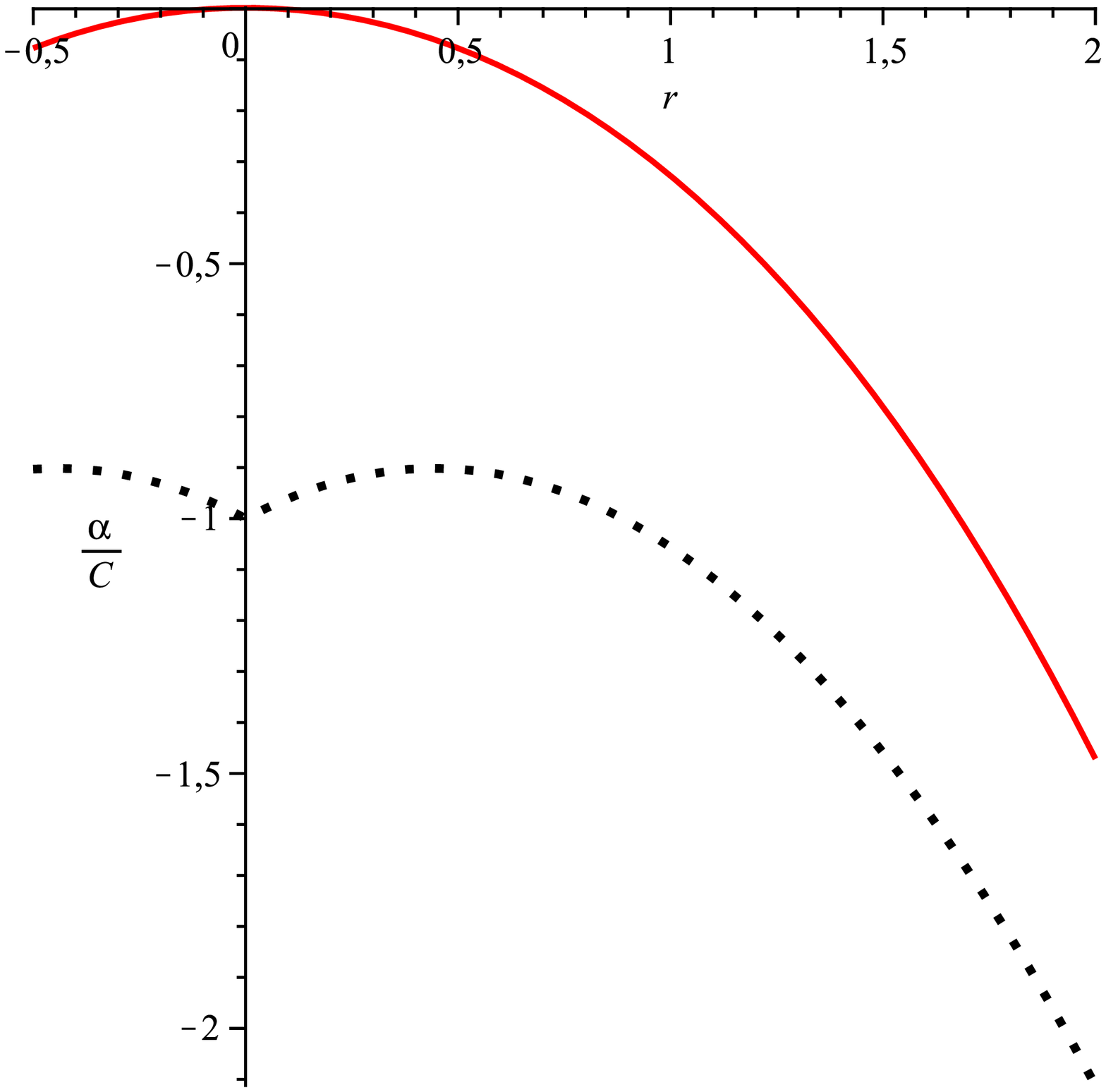}

      \caption{\label{figure3} 
      Behavior of $\alpha_\theta(x)$ for parameter values (from left):
   $q=0.5$, $m=0.55$; and $q=0.5$, $m=0.45$.  The dotted plot is the generic commutative solution for the parameters in question.      
}
     \end{center}
  \end{figure}

For $q\sim 1$, the cosmological parameter $|\Lambda|$ starts becoming more important with respect to  $\theta$.
In Fig. (\ref{figure2}), we see that as far $m=2.5$ the curve can rise, while for smaller values, i.e., $m=0.7$ the curve quickly drops down due to the influence of the cosmological term.  The interplay of $m$ and $q$ thus determines the overall causal nature of the spacetime: for large $m$ it is globally timelike, while for smaller $m$
the cosmological constant term produces large-scale spacelike regions. 
This effect is even more evident for $q=0.5$ (see Fig. \ref{figure3}). Depending on the value of the $m$, one has one or no horizon. As a special case when the curve intersects the $r$ axis at $r=0$, we should better speak of conical singularity rather than of a proper horizon. In any case we stress that there is no curvature singularity anywhere.

\begin{figure}
 \begin{center}
 \includegraphics[height=4.0cm]{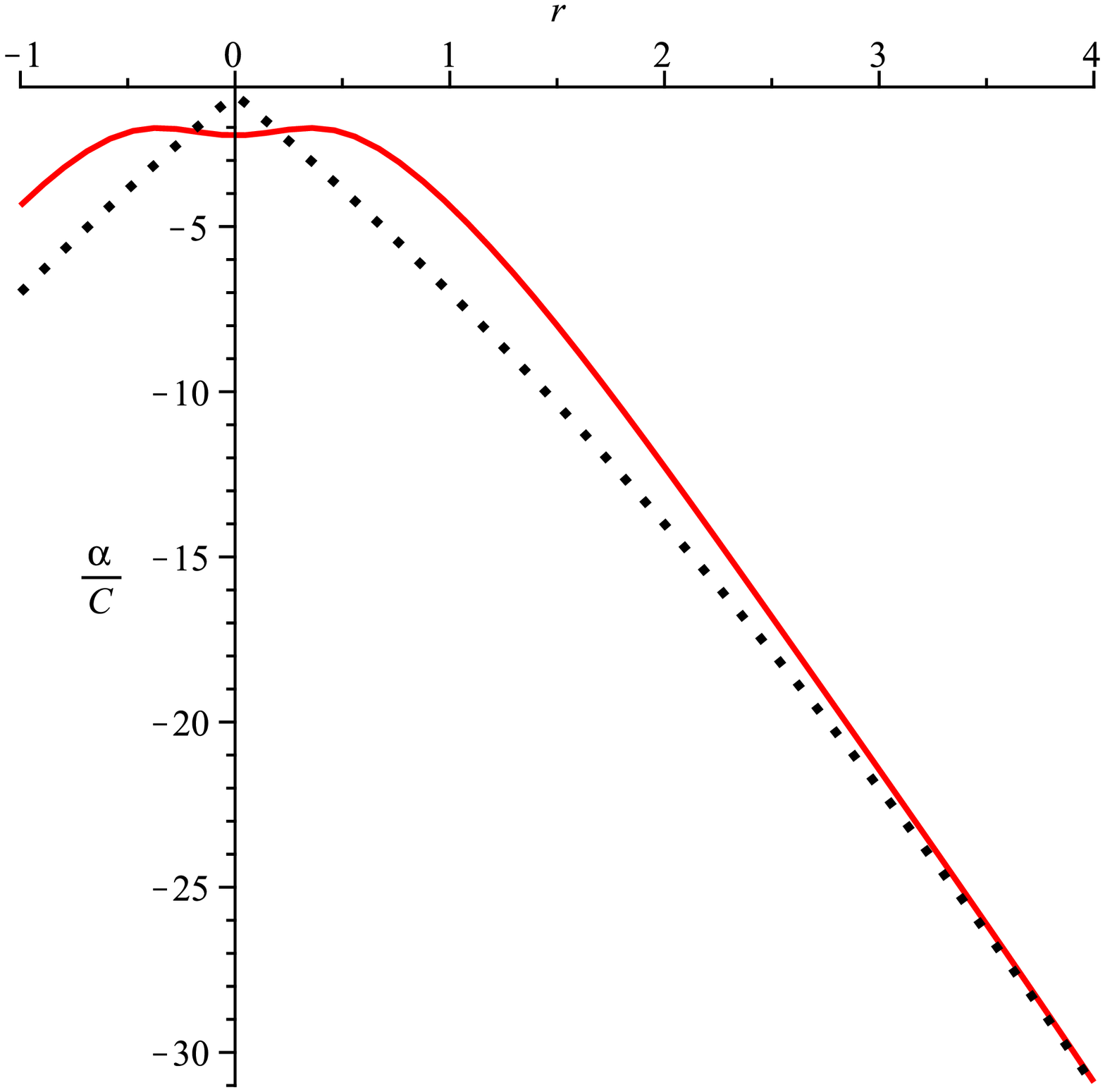}
  \includegraphics[height=4.0cm]{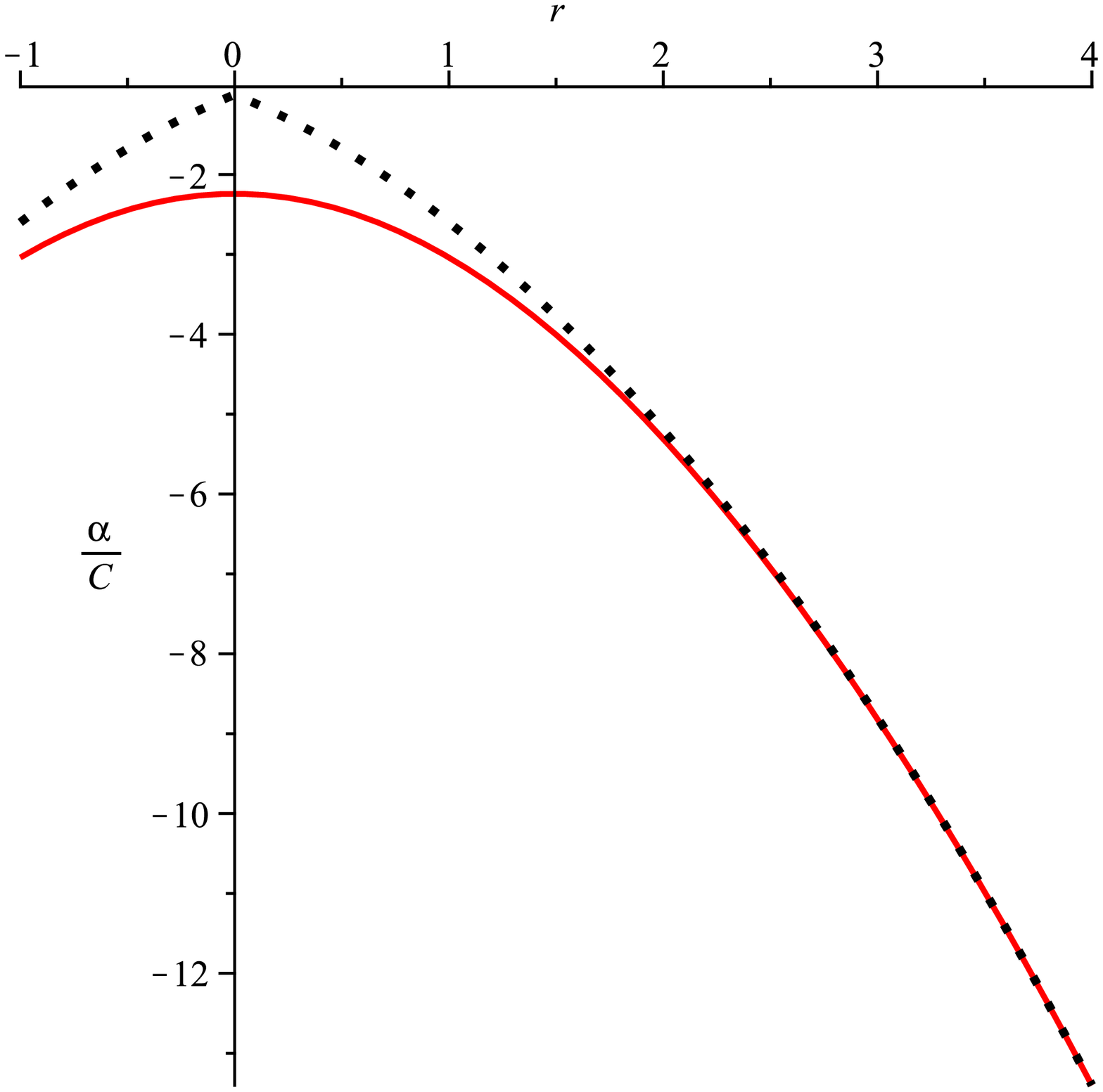}
 \includegraphics[height=4.0cm]{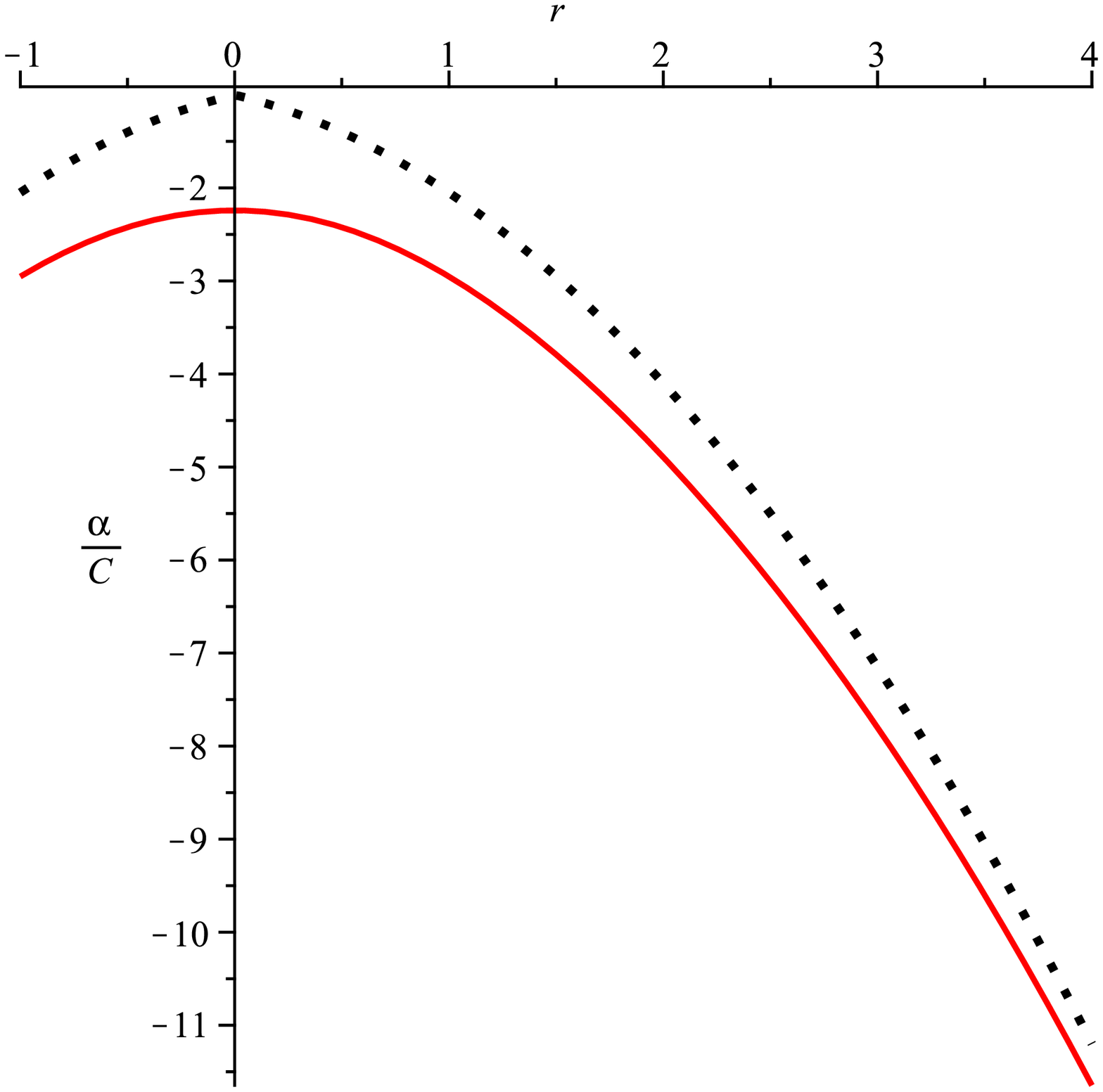}
 
      \caption{\label{figure4} 
      Behavior of $\alpha_\theta(x)$ for parameter values (from left):
   $q=5$, $m=-0.55$; $q=1$, $m=-0.55$; and $q=0.5$, $m=-0.55$.  The dotted plot is the generic commutative solution for the parameters in question.   
        }
     \end{center}
  \end{figure}

\begin{figure}
 \begin{center}
 \includegraphics[height=4.0cm]{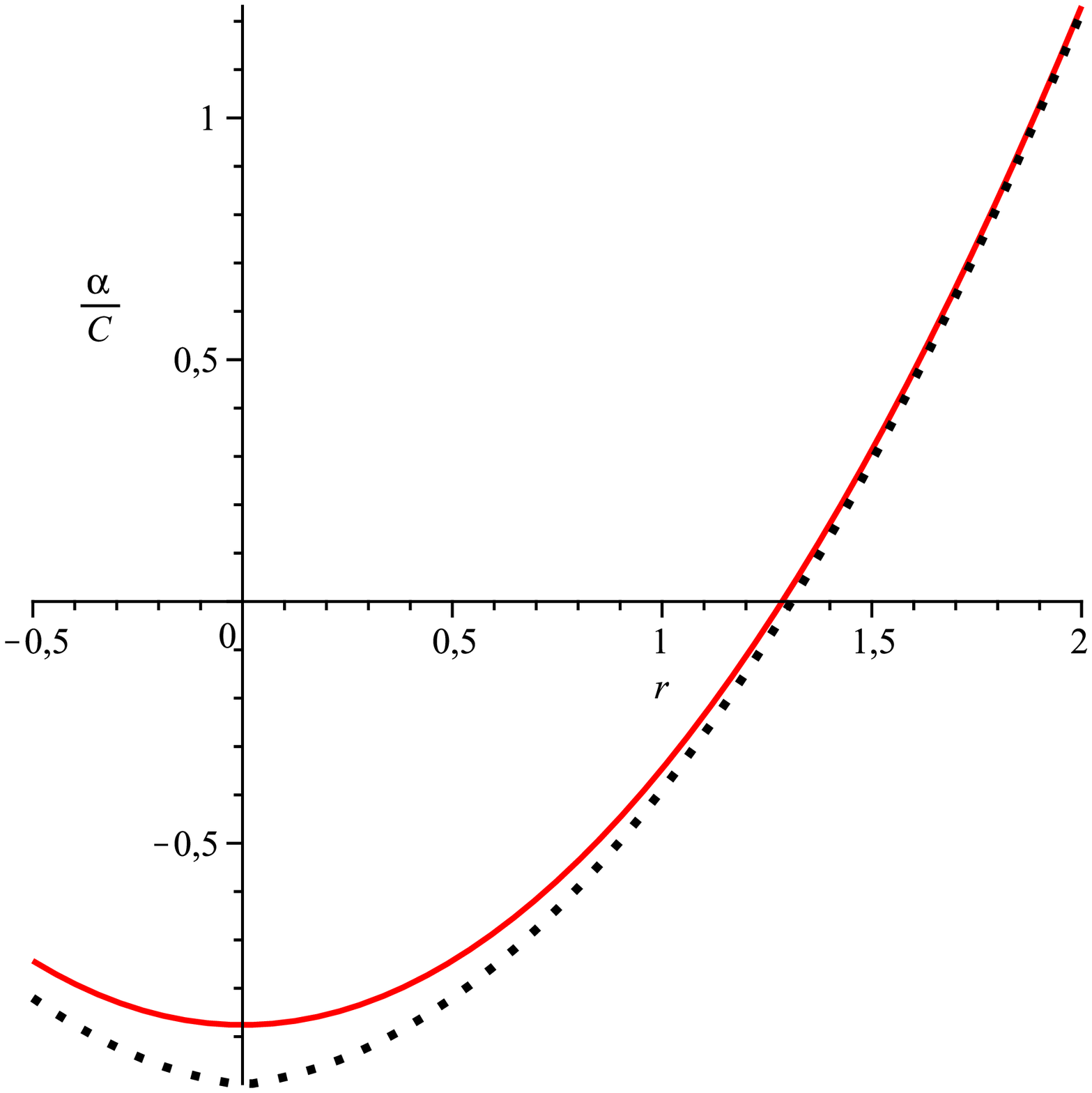}
  \includegraphics[height=4.0cm]{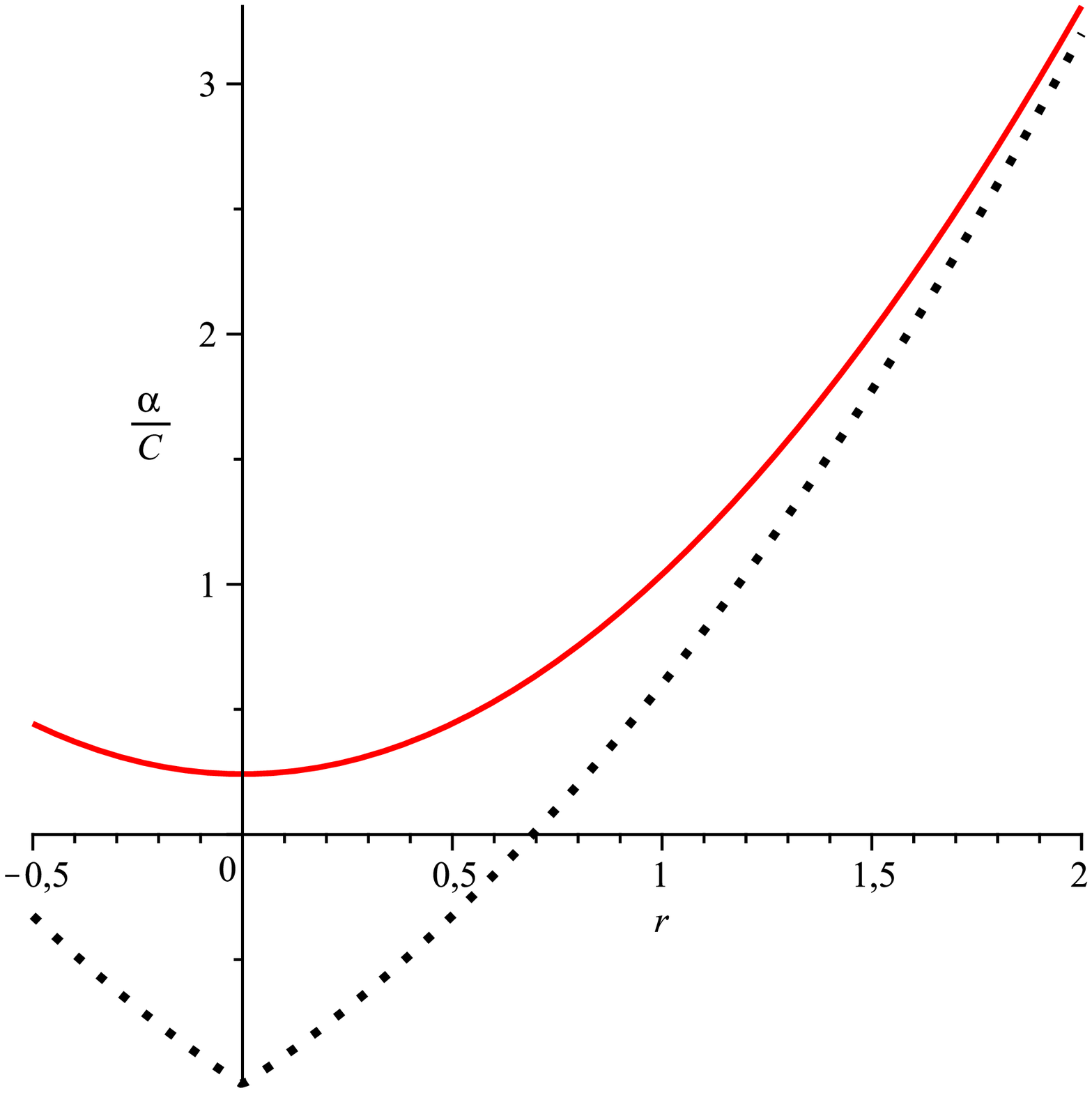}
 \includegraphics[height=4.0cm]{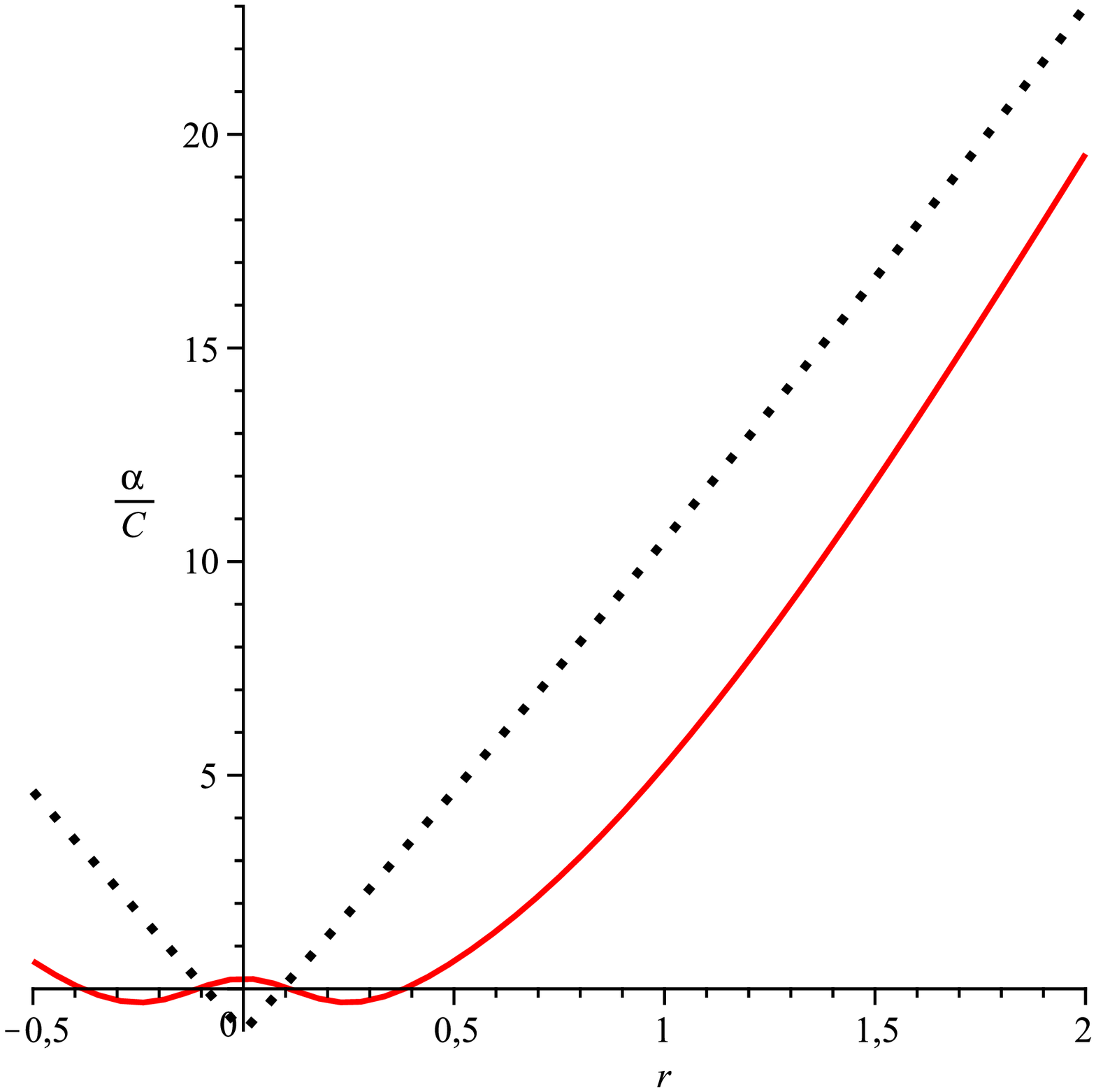}
 \includegraphics[height=4.0cm]{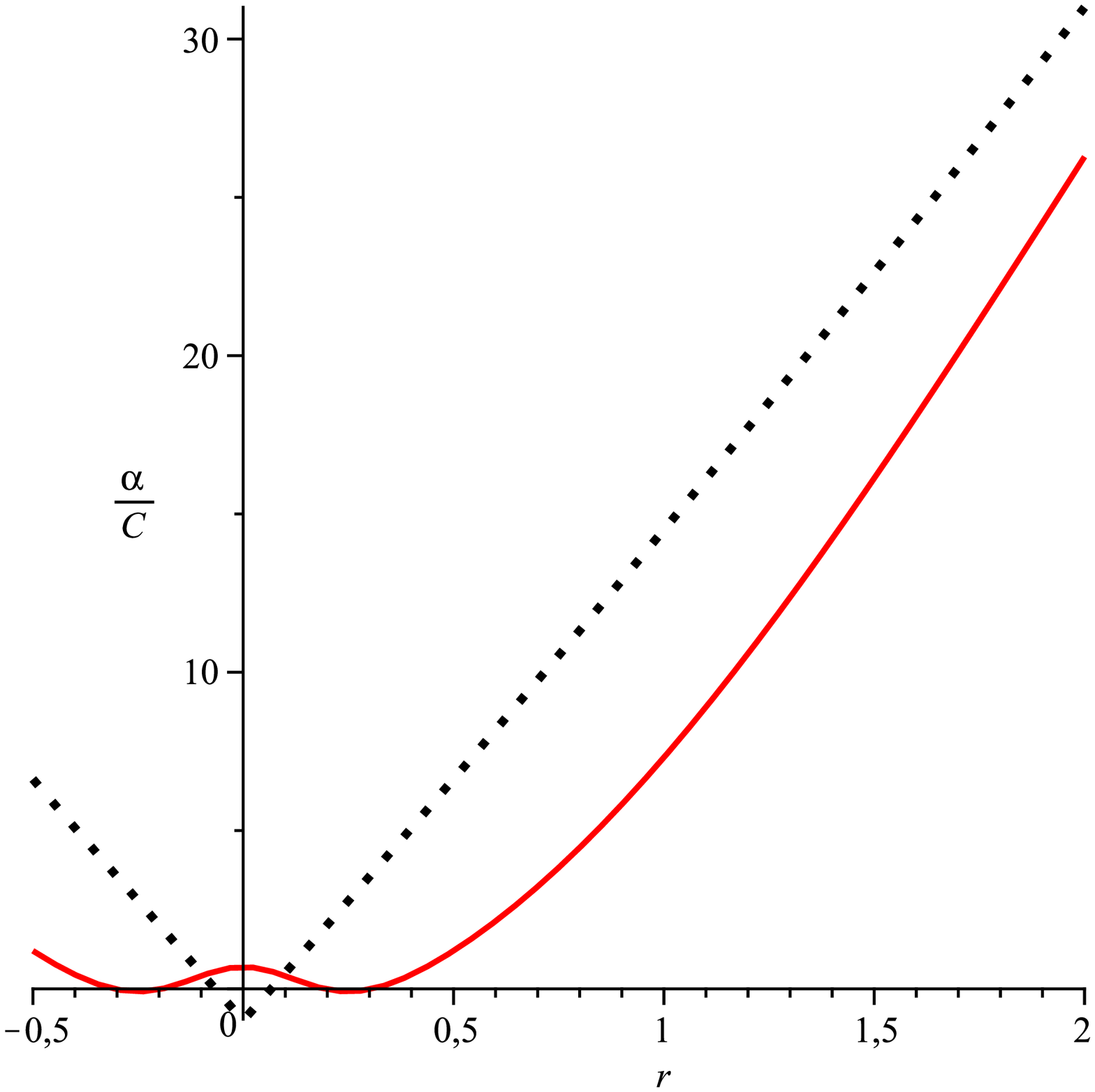}
 \includegraphics[height=4.0cm]{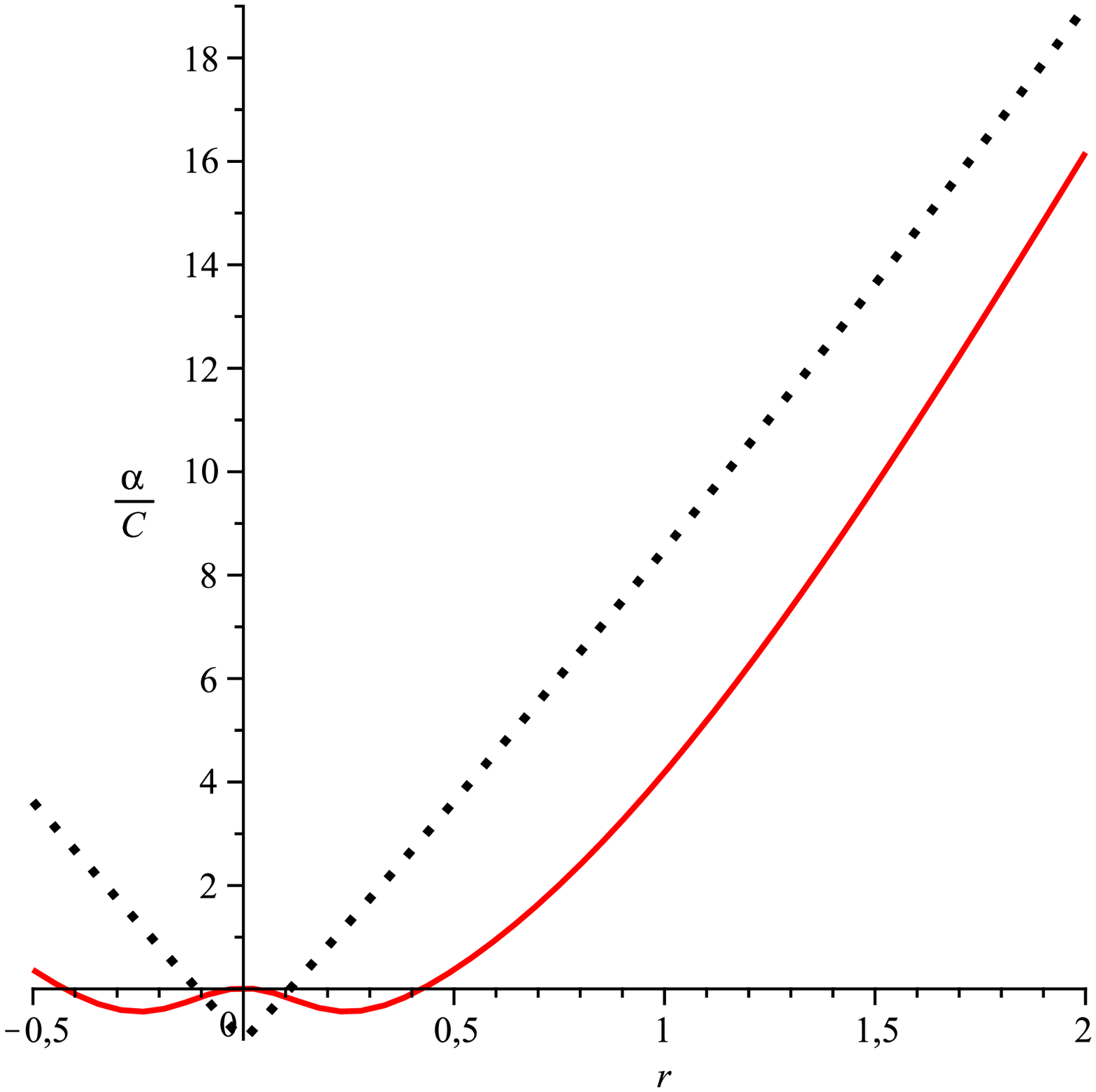}
 \includegraphics[height=4.0cm]{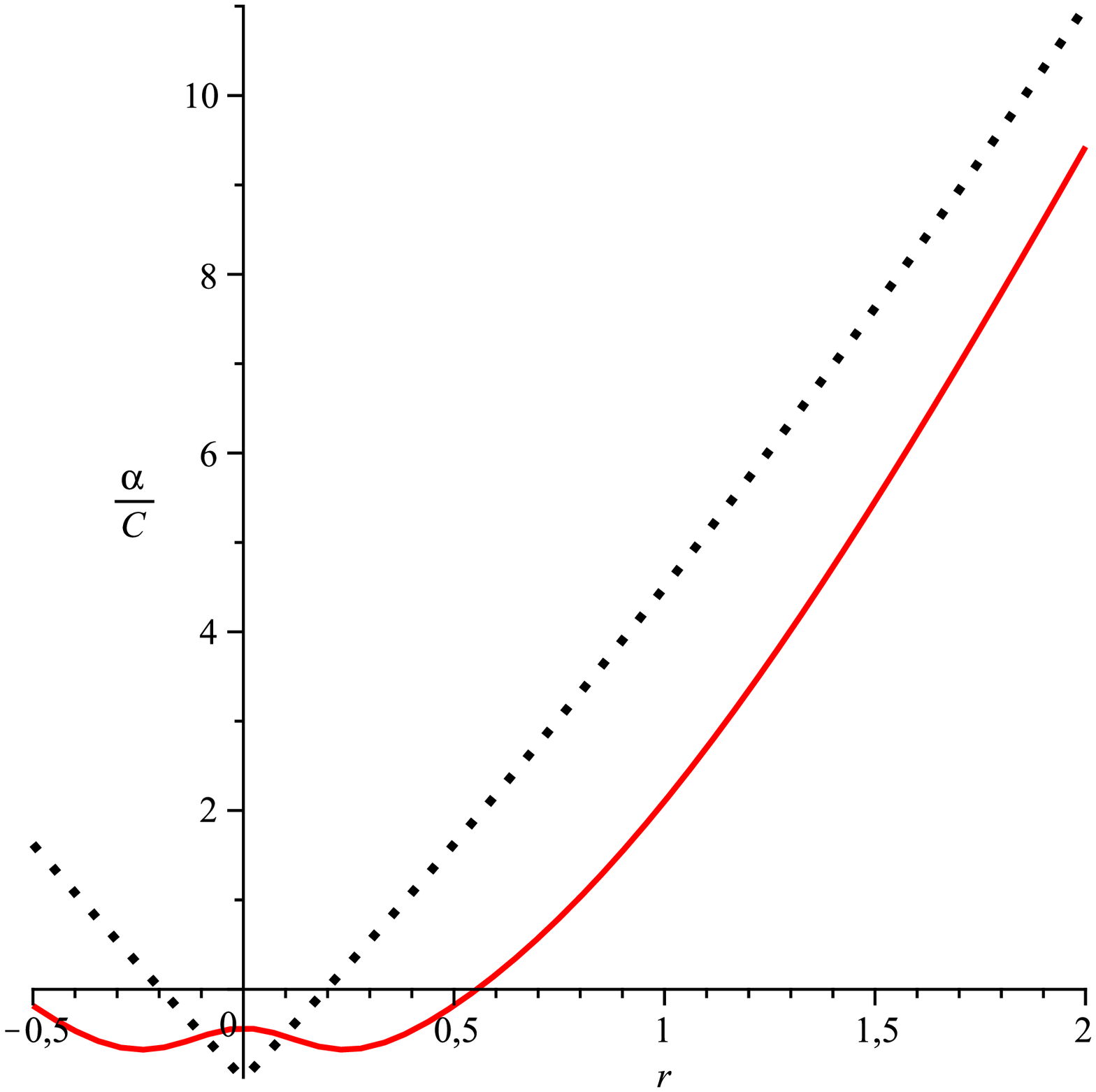}
 \includegraphics[height=4.0cm]{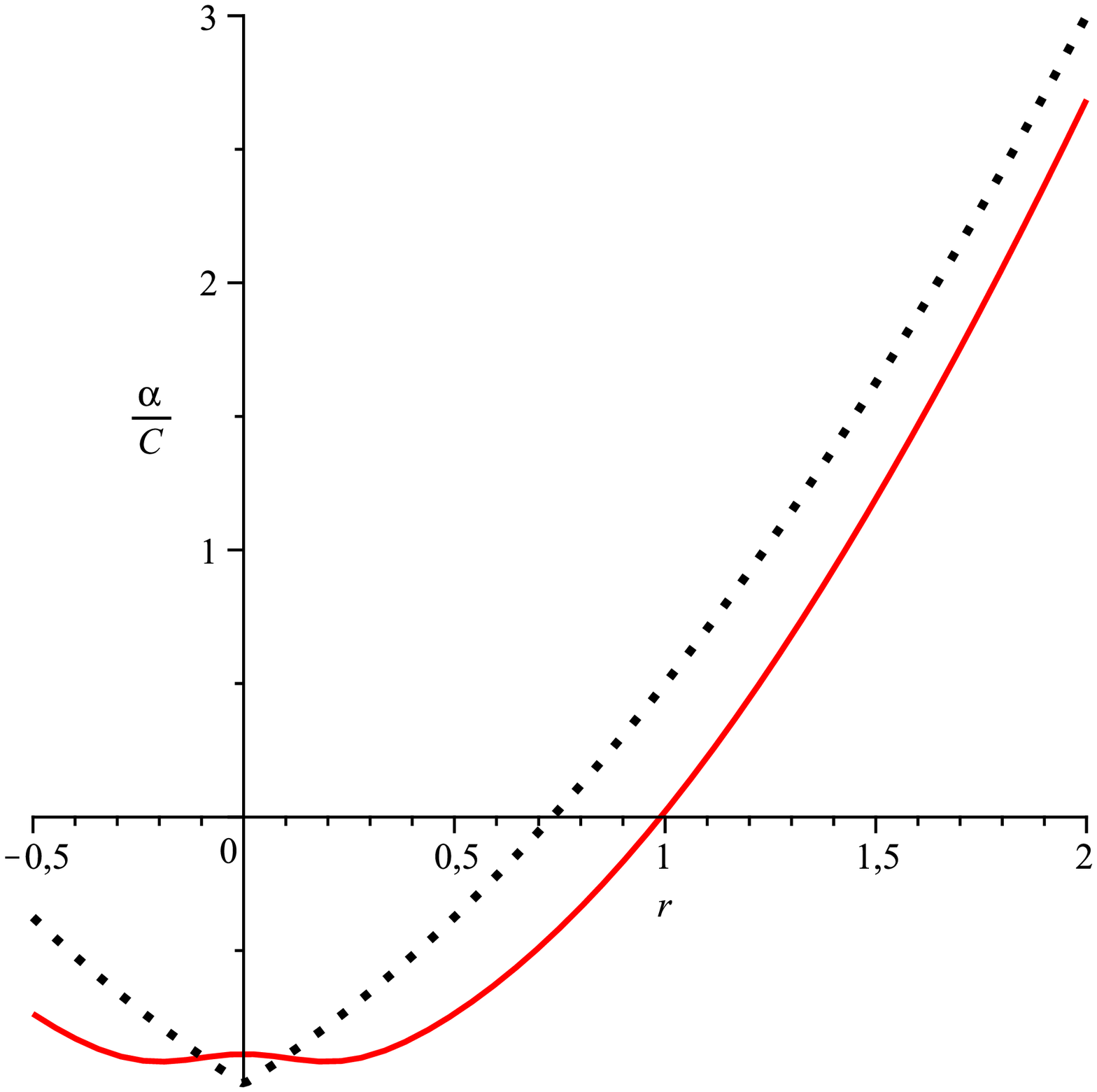}
 \includegraphics[height=4.0cm]{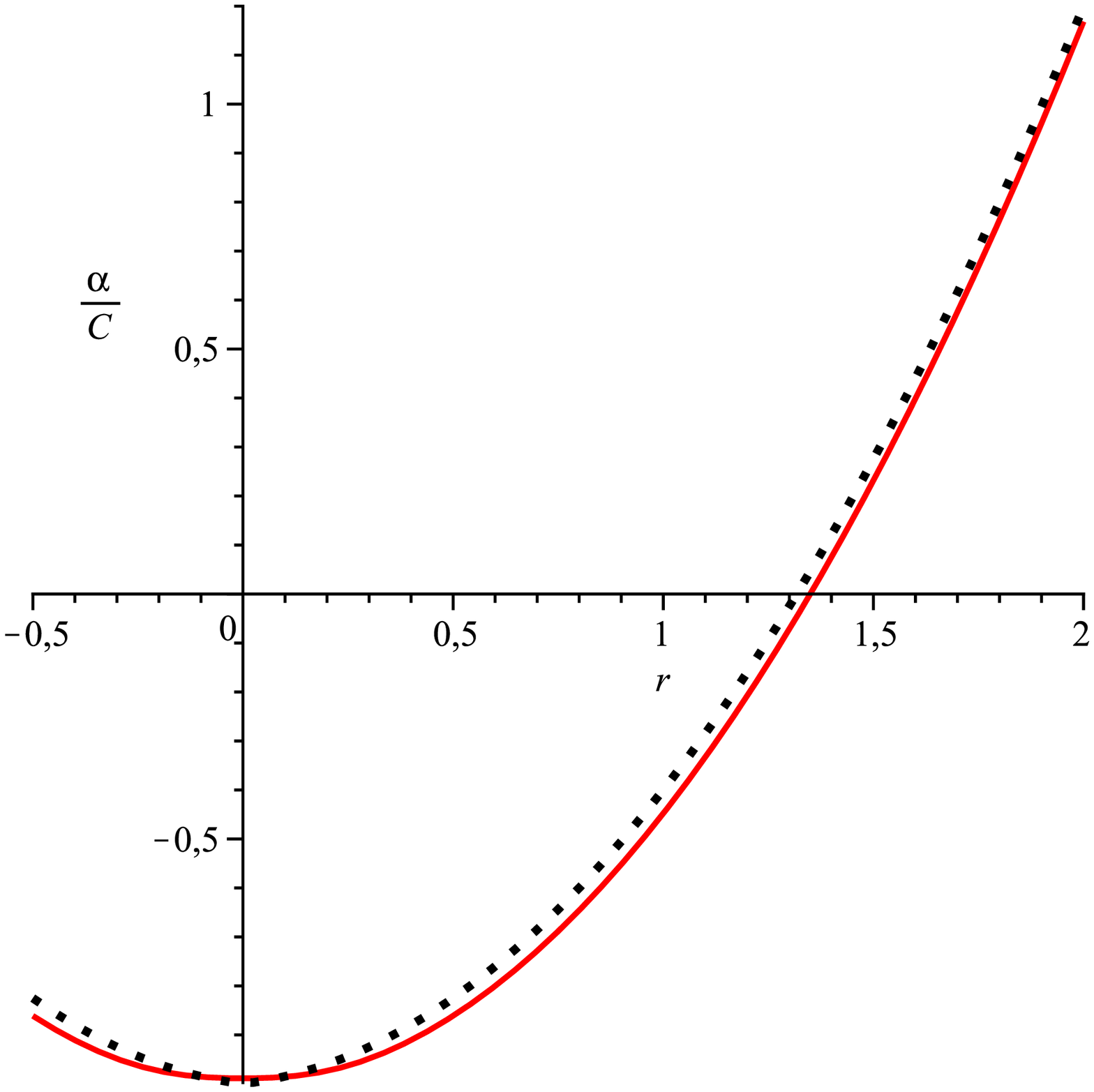}
 
      \caption{\label{figure5}
 Behavior of $\alpha_\theta(x)$ for parameter values (from left):
      $q=1$, $m=0.055$; $q=1$, $m=0.55$; $q=10$, $m=0.55$; $q=10$, $m=0.75$; $q=10$, $m=0.45$; $q=10$, $m=0.25$; $q=10$, $m=0.05$; and $q=10$, $m=0.005$.  The dotted plot is the generic commutative solution for the parameters in question.   
}
     \end{center}
  \end{figure}

\begin{figure}
 \begin{center}
 \includegraphics[height=4.0cm]{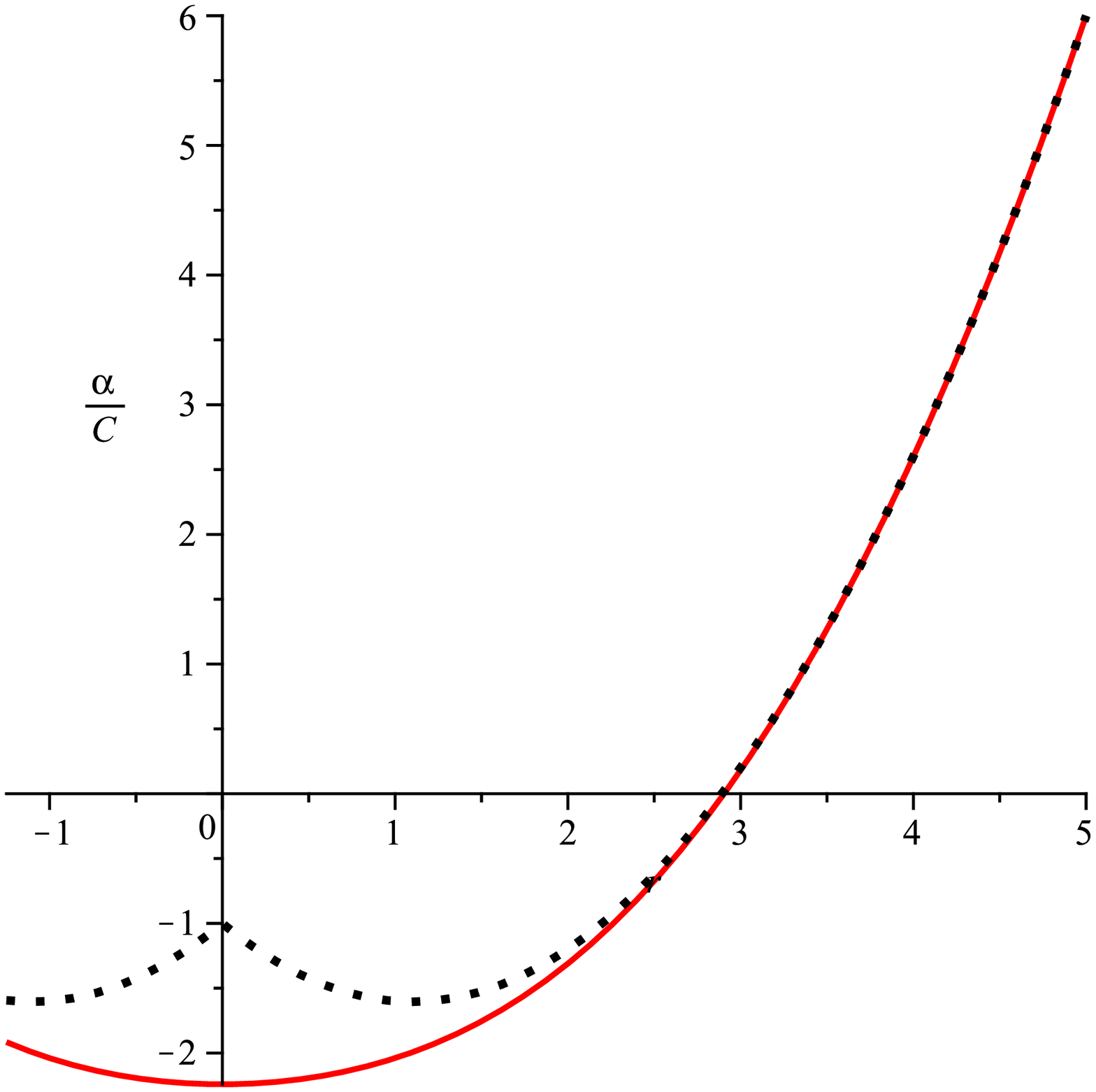}
 \includegraphics[height=4.0cm]{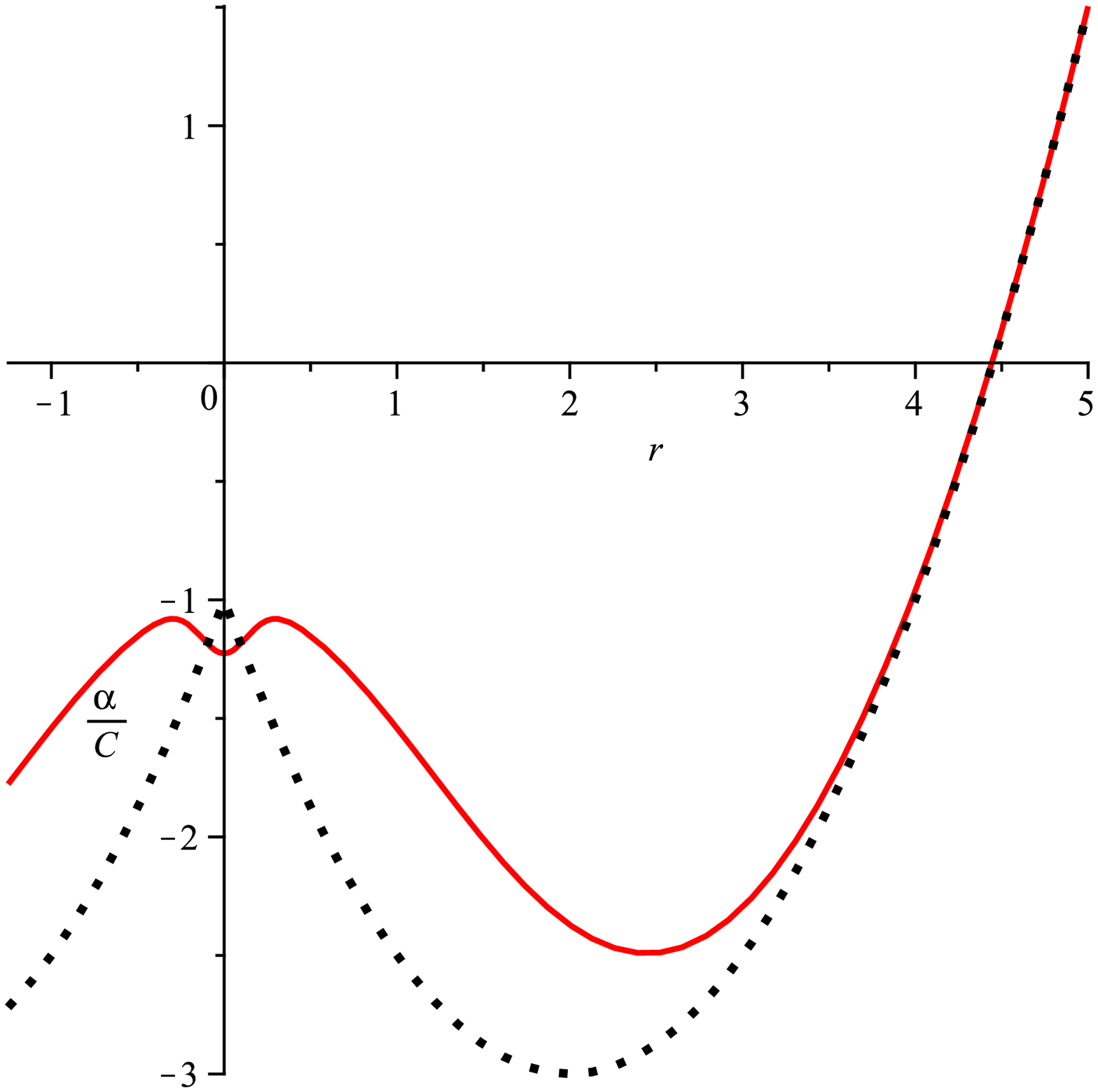}
 
      \caption{\label{figure6}
Behavior of $\alpha_\theta(x)$ for parameter values (from left):
      $q=1$, $m=-0.55$; $q=10$, $m=-0.1$.  The dotted plot is the generic commutative solution for the parameters in question.   
      }
     \end{center}
  \end{figure}

The metric admits three further combinations of signs for $\Lambda/C$ and $m$ which give rise to distinct spacetime geometries. For negative $m$ (and $\Lambda/C>0$), the metric does not have any horizon. Indeed starting from the value $-1-4|m|/\sqrt{\pi}$ at $r=0$, the curve keeps decreasing to further negative values (see Fig. \ref{figure4}). For negative $\Lambda/C$ (and $m>0$), the curve starts at $r=0$ with a value $-1+4|m|/\sqrt{\pi}$, which can be either positive of negative, depending on the value of $m$. For small $q$ the function $\alpha_\theta (r)/C$ is monotonically increasing. Only the exponential has a negative derivative which is suppressed for small $q$. Thus, there will be one or no horizon according to the negative of positive value at the origin. However for large $q$ the exponential term quickly dies off producing a local decrease of the curve. As a result there may be two horizons or one horizon and a conical singularity at the origin or just a single degenerate horizon (see Fig. \ref{figure5}). As we shall see later when studying the temperature the extremal configuration for $m\approx 0.75$ actually does not occur during the evaporation. This is simply the bigger value of $m$ to have horizons. For both negative $\Lambda/C$ and $m$, we have that the curve starts at $r=0$ with a negative value $-1-4|m|/\sqrt{\pi}$. For small $q$ the curve grows and one horizon occurs. For large $q$ the curve can locally grow but can reach the $r$ axis only one time (see Fig. \ref{figure6}).

\begin{figure}
 \begin{center}
 \includegraphics[height=4.0cm]{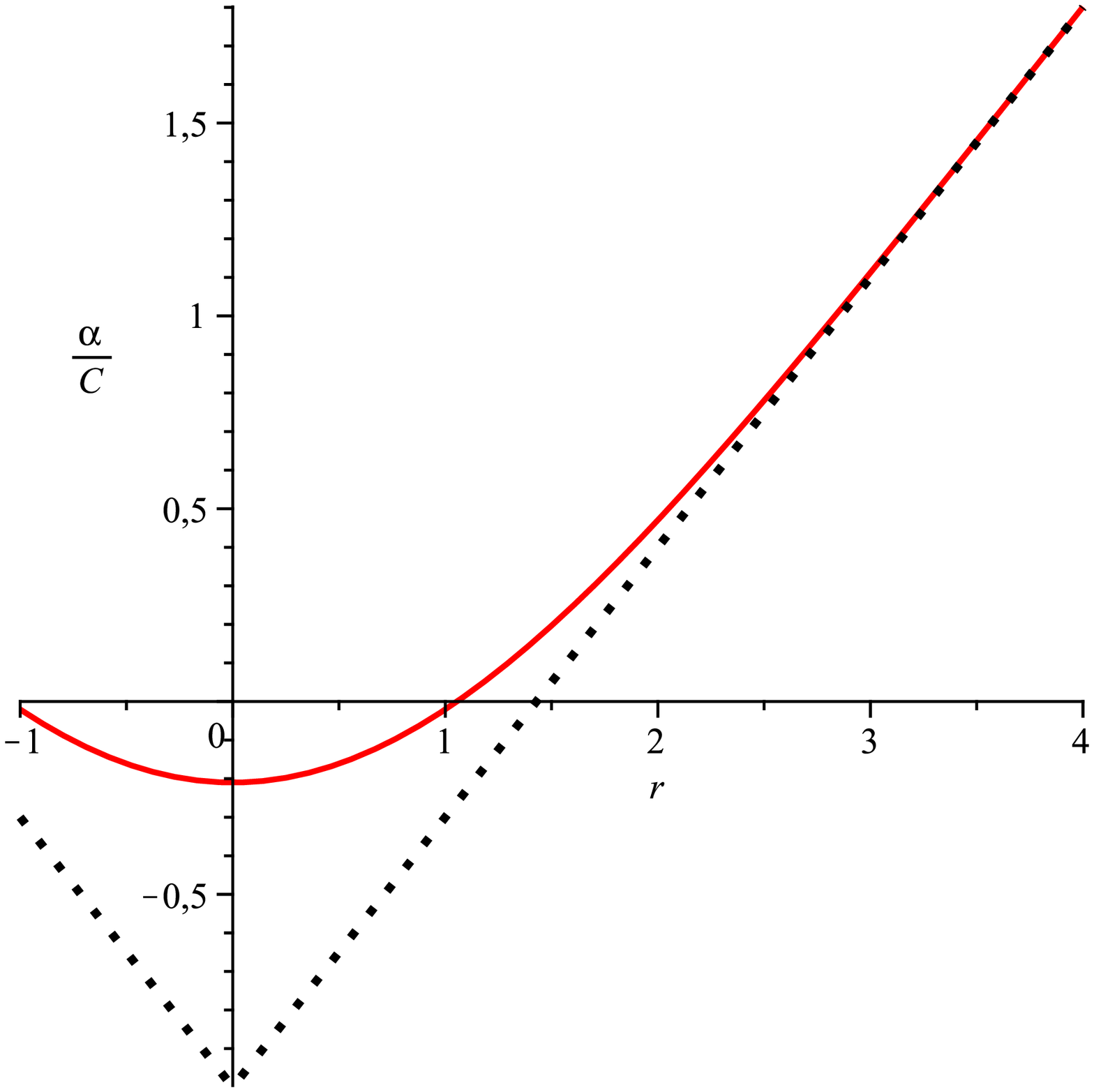}
 \includegraphics[height=4.0cm]{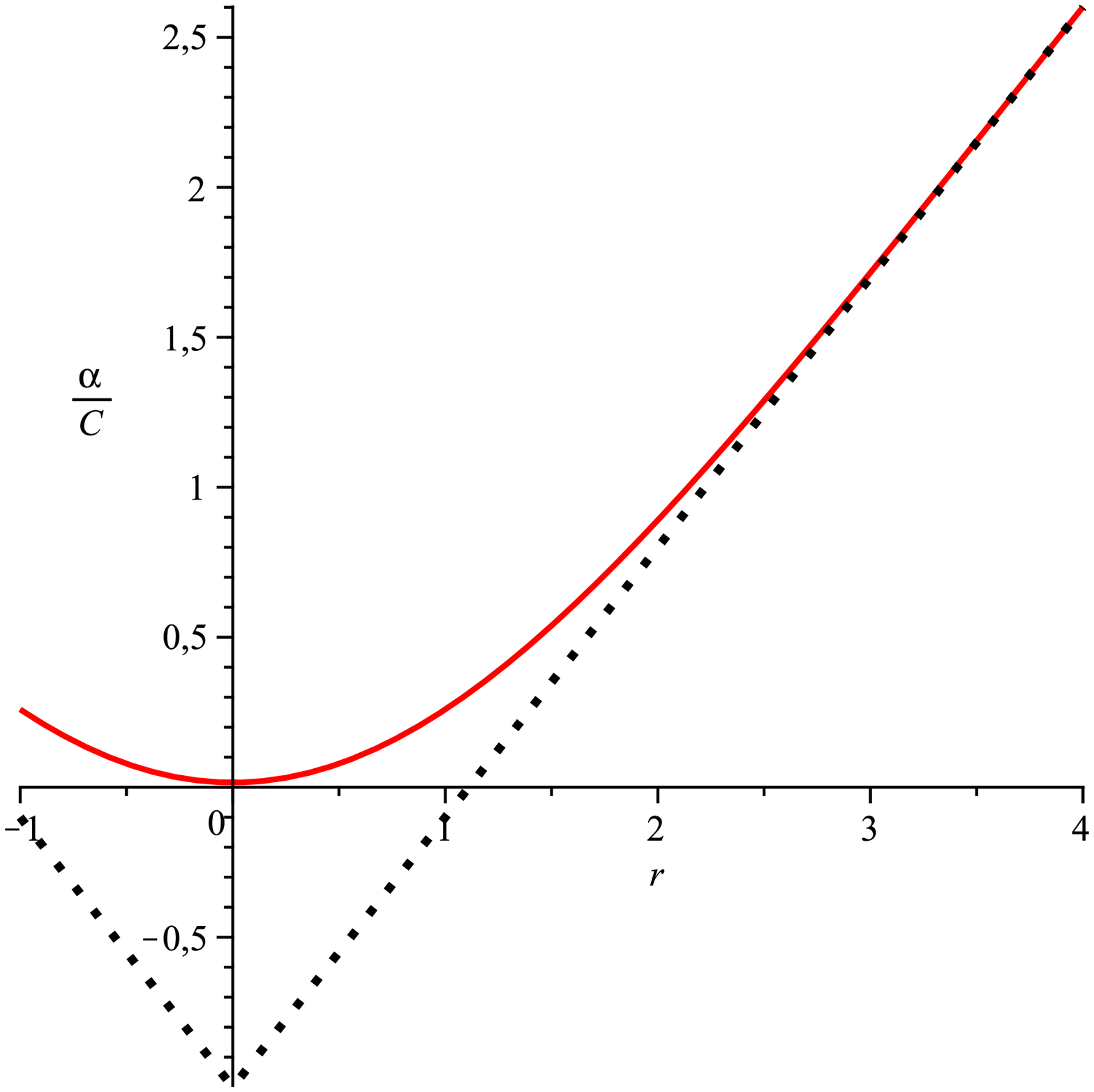}
 \includegraphics[height=4.0cm]{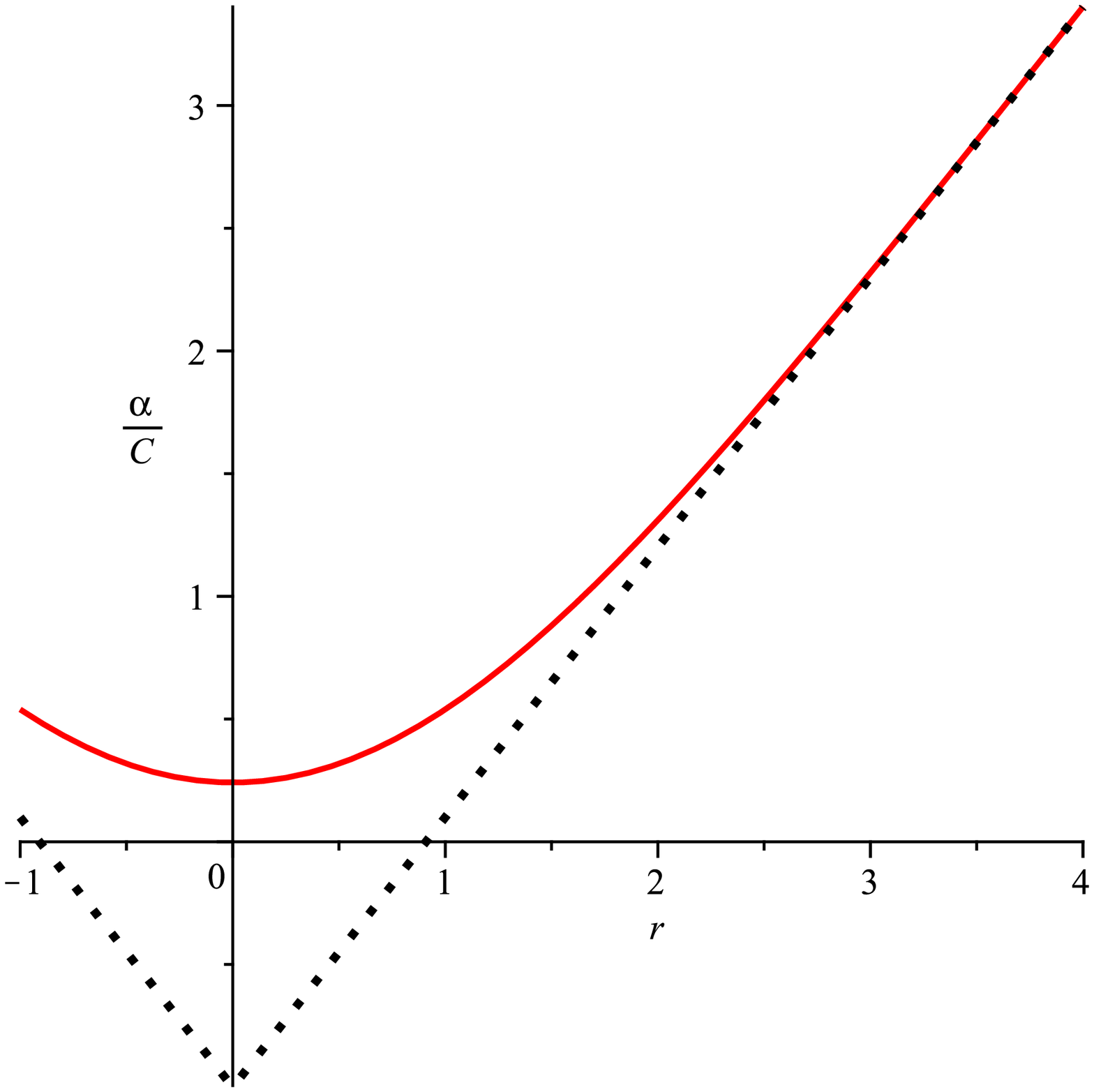}
 \includegraphics[height=4.0cm]{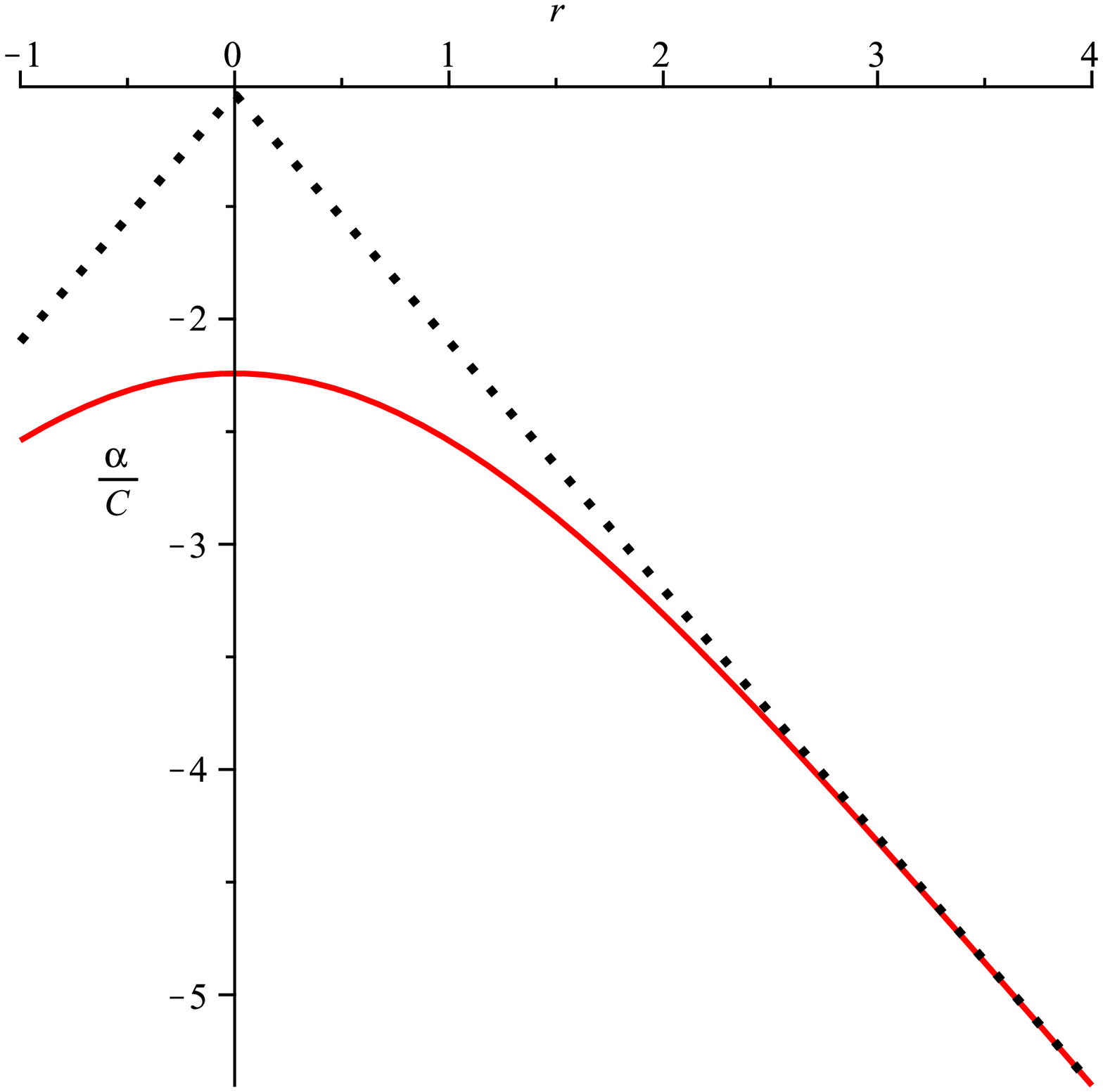}
 
      \caption{\label{figure7}
Behavior of $\alpha_\theta(x)$ for parameter values (from left):
      $m=0.35$, $m=0.45$, $m=0.55$ and $m=-0.55$.  The dotted plot is the generic commutative solution for the parameters in question.   
     }
     \end{center}
  \end{figure}

Our analysis proceeds with the case of vanishing $\Lambda$. We can write 
\begin{equation}
\frac{\alpha_\theta(r)}{C}=-1+\frac{4m}{\sqrt{\pi}} \left[ \exp\left(-\frac{ r^2}{4}\right)
+\frac{1}{2} |r|  \gamma\left(\half,\frac{r^2}{4}\right)\right]
\end{equation}
where $m=M\sqrt{\theta}/C$ and $r=x/\sqrt{\theta}$. For positive $m$, the curve starts at $r=0$ with a value $-1+4m/\sqrt{\pi}$, which can be positive, negative or vanishing depending on the value of $m$. Then the curve grows giving rise to one or no horizon or eventually at a conical singularity at the origin. Conversely for negative $m$ there is no horizon since the curve starts at a negative value and keeps decreasing (see Fig. \ref{figure7}).

\begin{figure}
 \begin{center}
 \includegraphics[height=4.0cm]{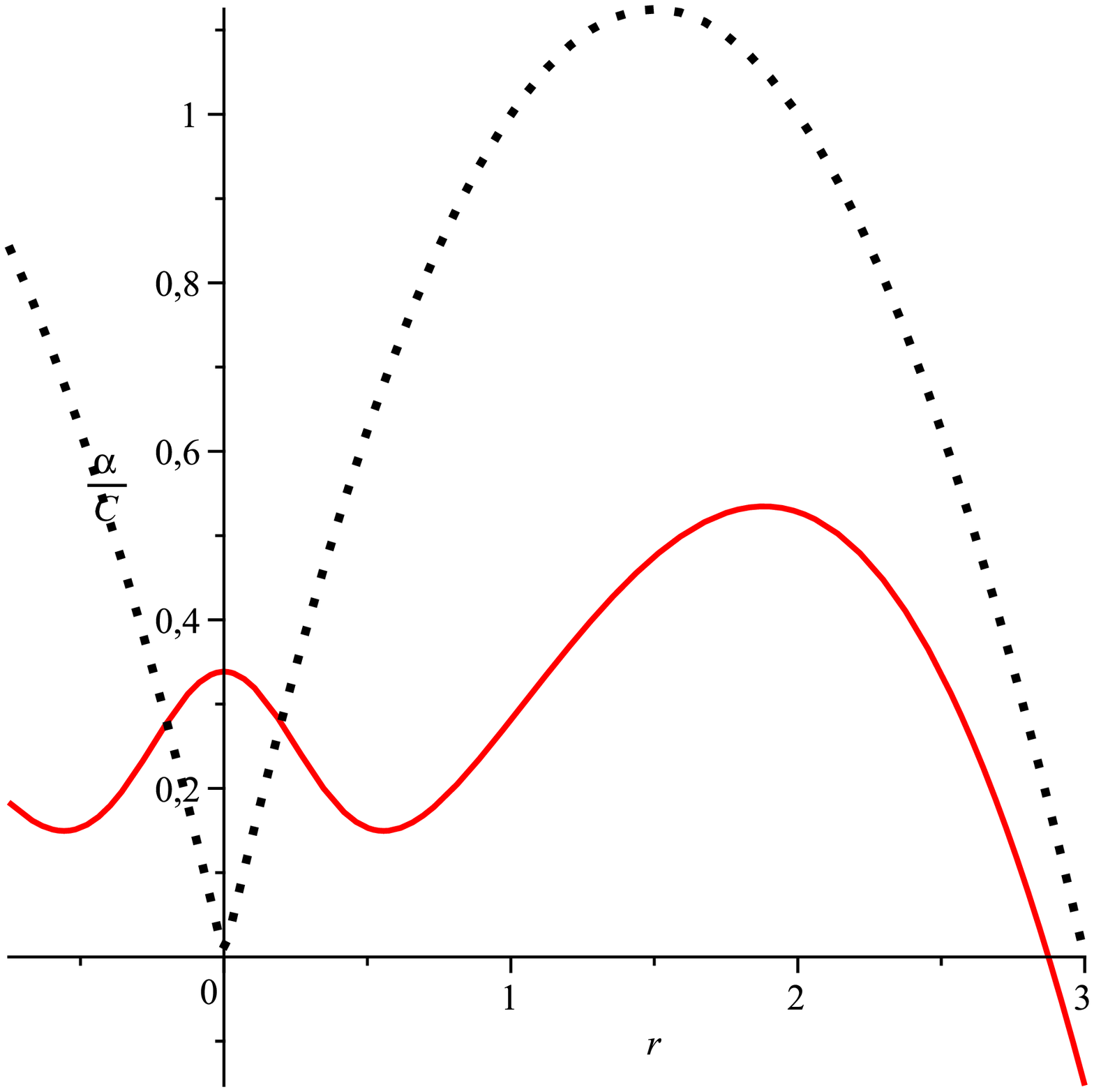}
 \includegraphics[height=4.0cm]{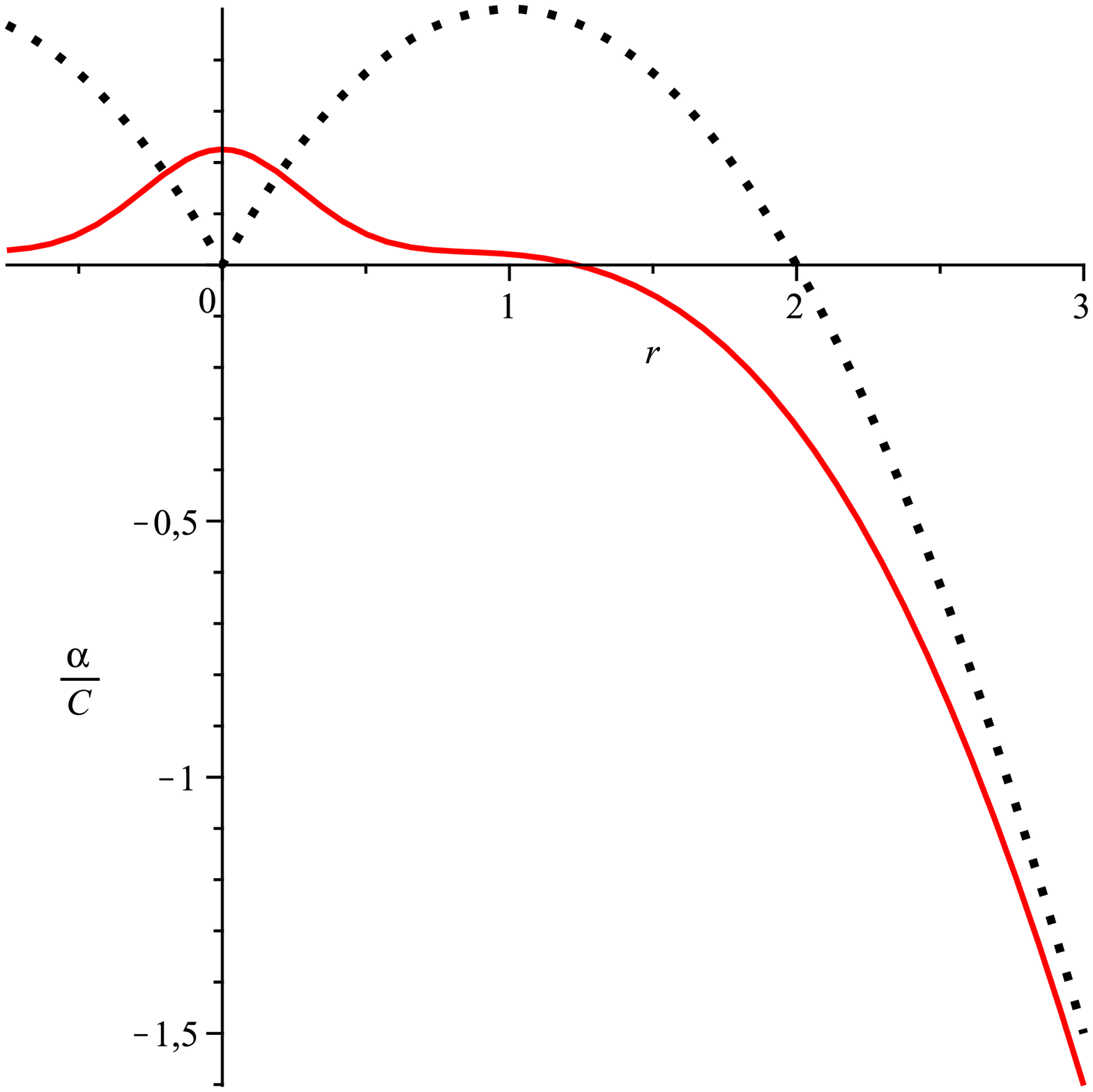}
 \includegraphics[height=4.0cm]{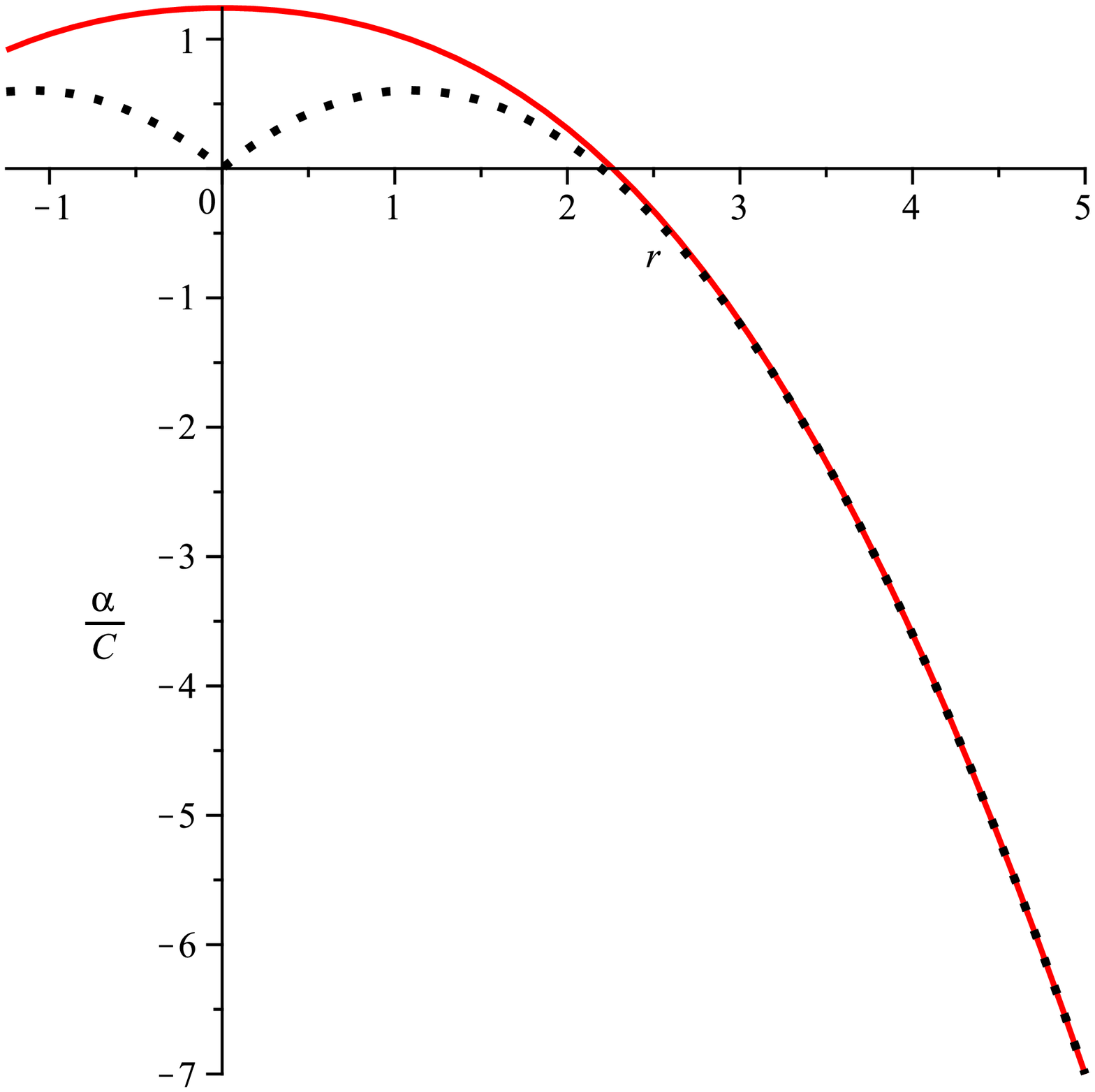}
 
      \caption{\label{figure8}
Behavior of $\alpha_\theta(x)$ for parameter values (from left):
      $q=5$, $m=0.15$; $q=5$, $m=0.1$; $q=1$, $m=0.55$.  The dotted plot is the generic commutative solution for the parameters in question.   
            }
     \end{center}
  \end{figure}

\begin{figure}
 \begin{center}
 \includegraphics[height=4.0cm]{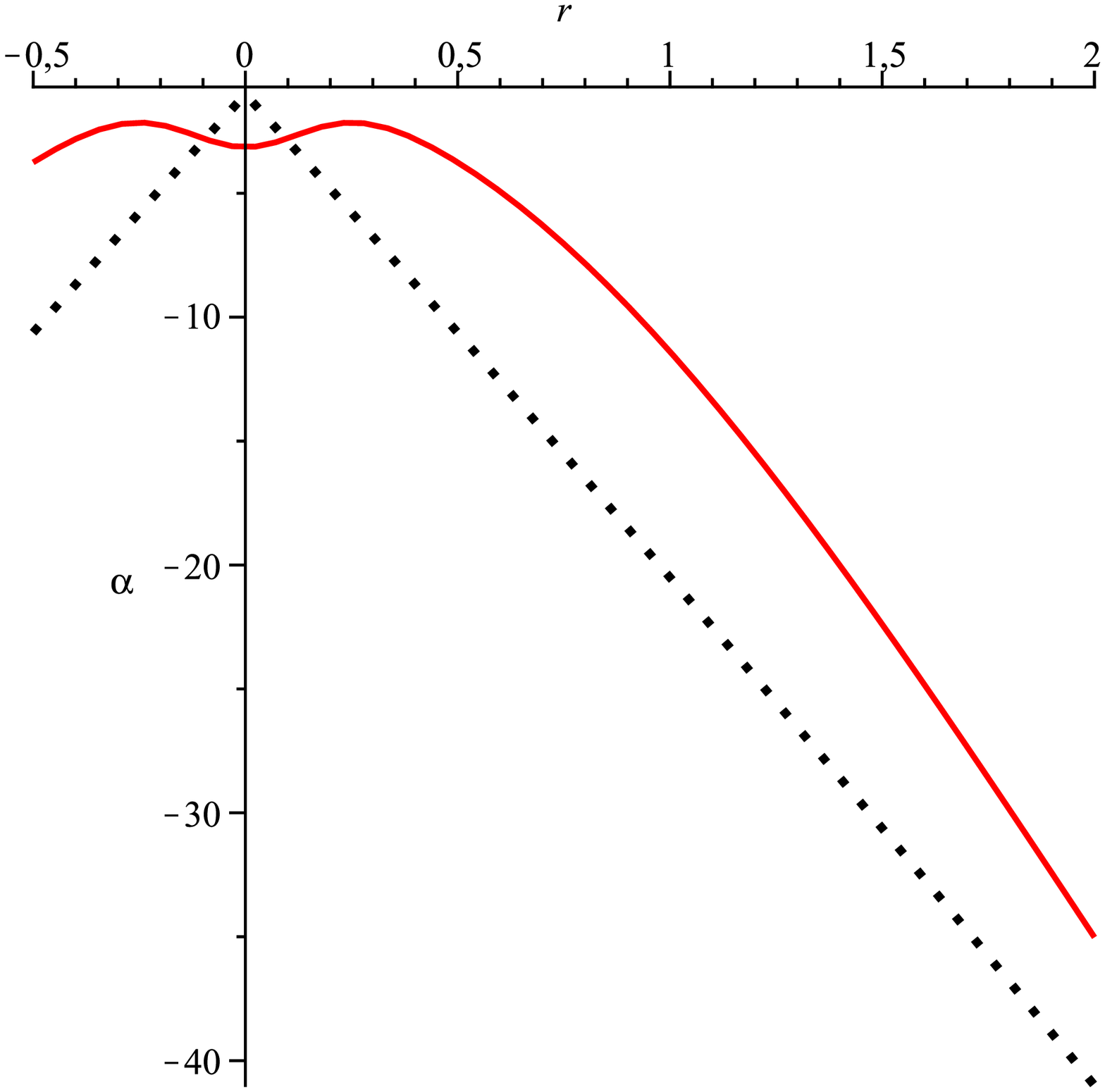}
 \includegraphics[height=4.0cm]{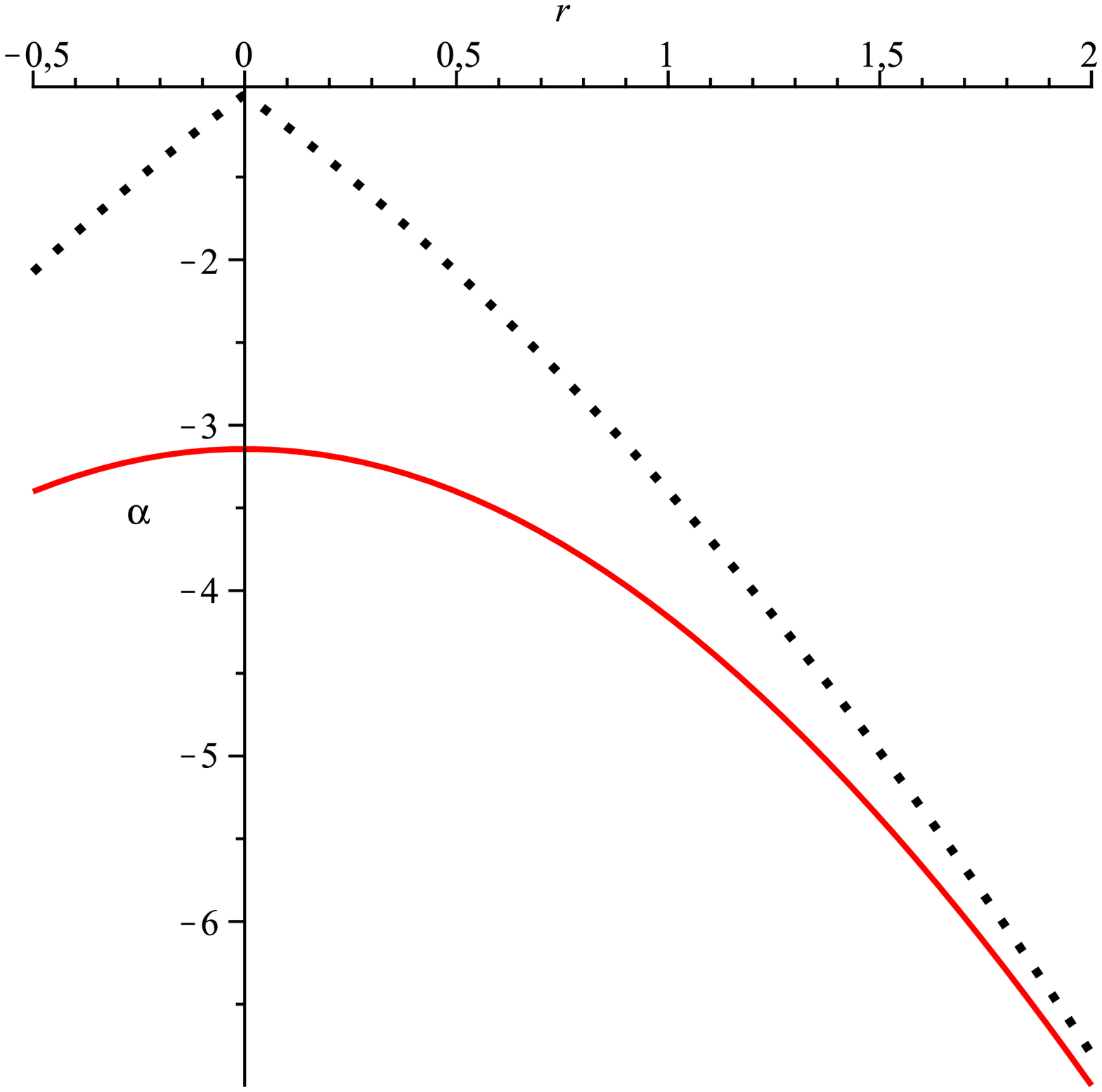}
 
      \caption{\label{figure9}
      Behavior of $\alpha_\theta(x)$ for parameter values (from left):
      $q=10$, $m=-0.95$; and $q=1$, $m=-0.95$.  The dotted plot is the generic commutative solution for the parameters in question.   
      }
     \end{center}
  \end{figure}

For vanishing $C$ we can write
\begin{equation}
\alpha_\theta(r)=\mp\frac{1}{2}\ r^2+\frac{4m}{\sqrt{\pi}} \left[ \exp\left(-\frac{q^2 r^2}{4}\right)
+\frac{1}{2} (q|r|)  \gamma\left(\half,\frac{q^2r^2}{4}\right)\right]
\end{equation}
where $r=\sqrt{|\Lambda|}\ x$, $m=M\sqrt{\theta}$ and $q =1/\sqrt{|\Lambda|\theta}$. We notice that the line element is symmetric for change of sign of both $\Lambda$ and $m$. Thus we need to study two cases.
For both positive $m$ and $\Lambda$ and $q=5$, the curve starts at $r=0$ with a positive value $4m/\sqrt{\pi}$, the descends due to the exponential term at short scales, grows up for the term containing the $\gamma$ and dies off quadratically at large distances due to the cosmological term $-\half r^2$.  If we have $q=1$, the $\gamma$ term is negligible and the curves rapidly lowers down provided $m$ is small. In any case we find just one horizon (see Fig. \ref{figure8}). For negative $m$ the curve starts with a negative value at the origin. After a small increase due to the exponential term, it lowers down without intersection the $r$ axis. Therefore there is no horizon (see Fig. \ref{figure9}).

The case $\Lambda=C=0$ corresponds to a regular geometry in which no horizon occurs.

\section{Thermodynamics}

  \begin{figure}
 \begin{center}
 \includegraphics[height=3.7cm]{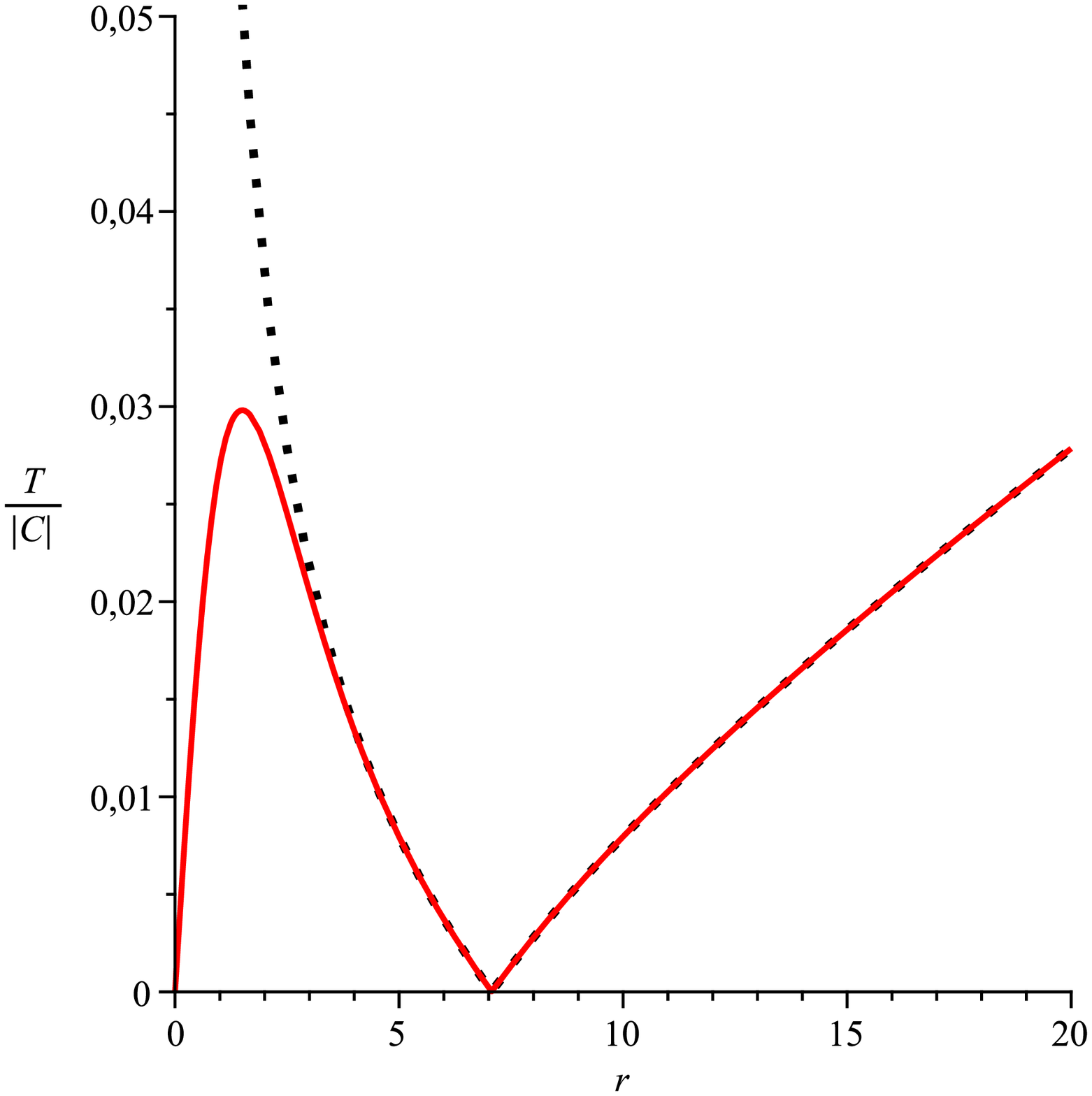}
  \includegraphics[height=3.7cm]{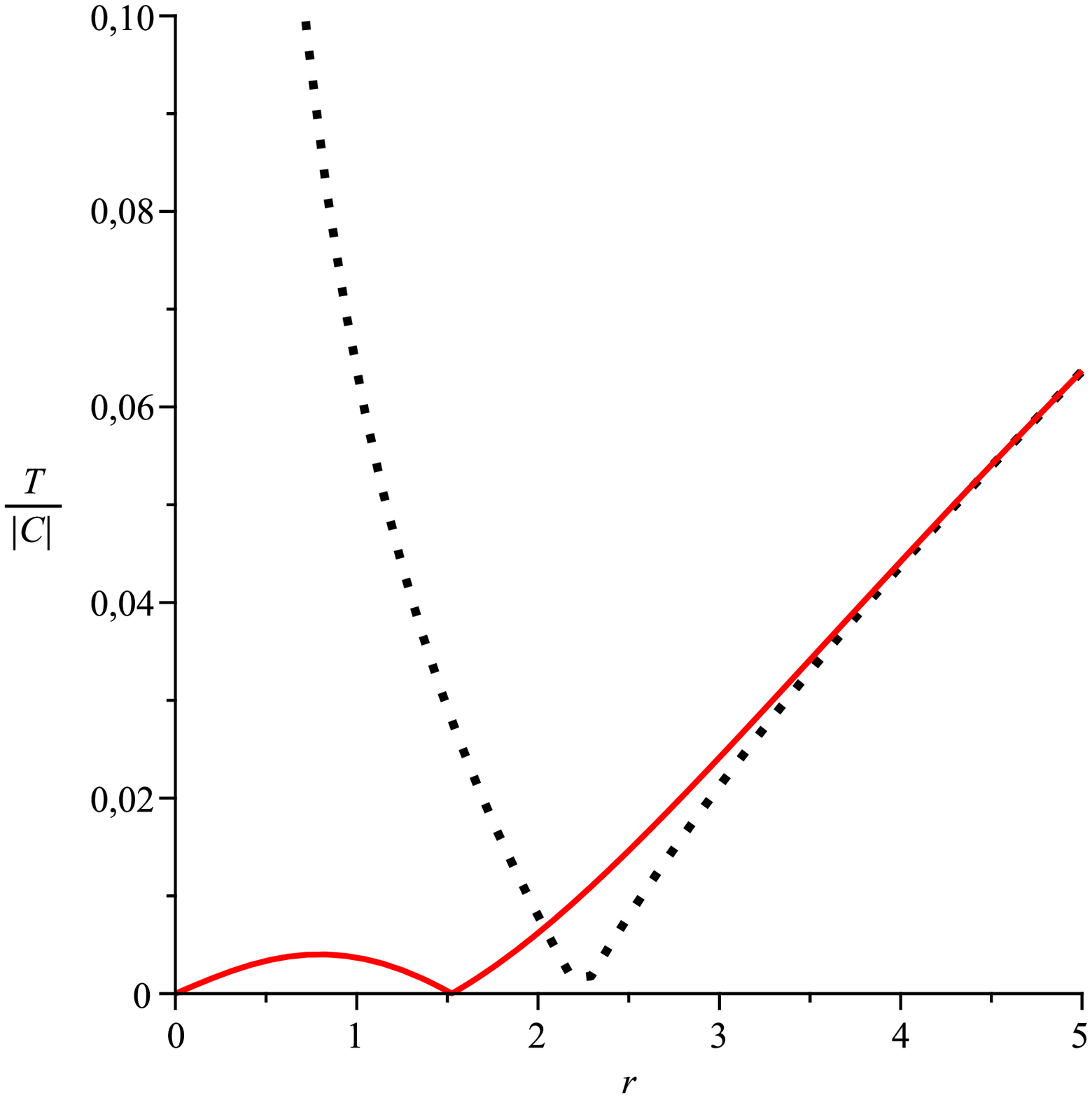}
 \includegraphics[height=3.7cm]{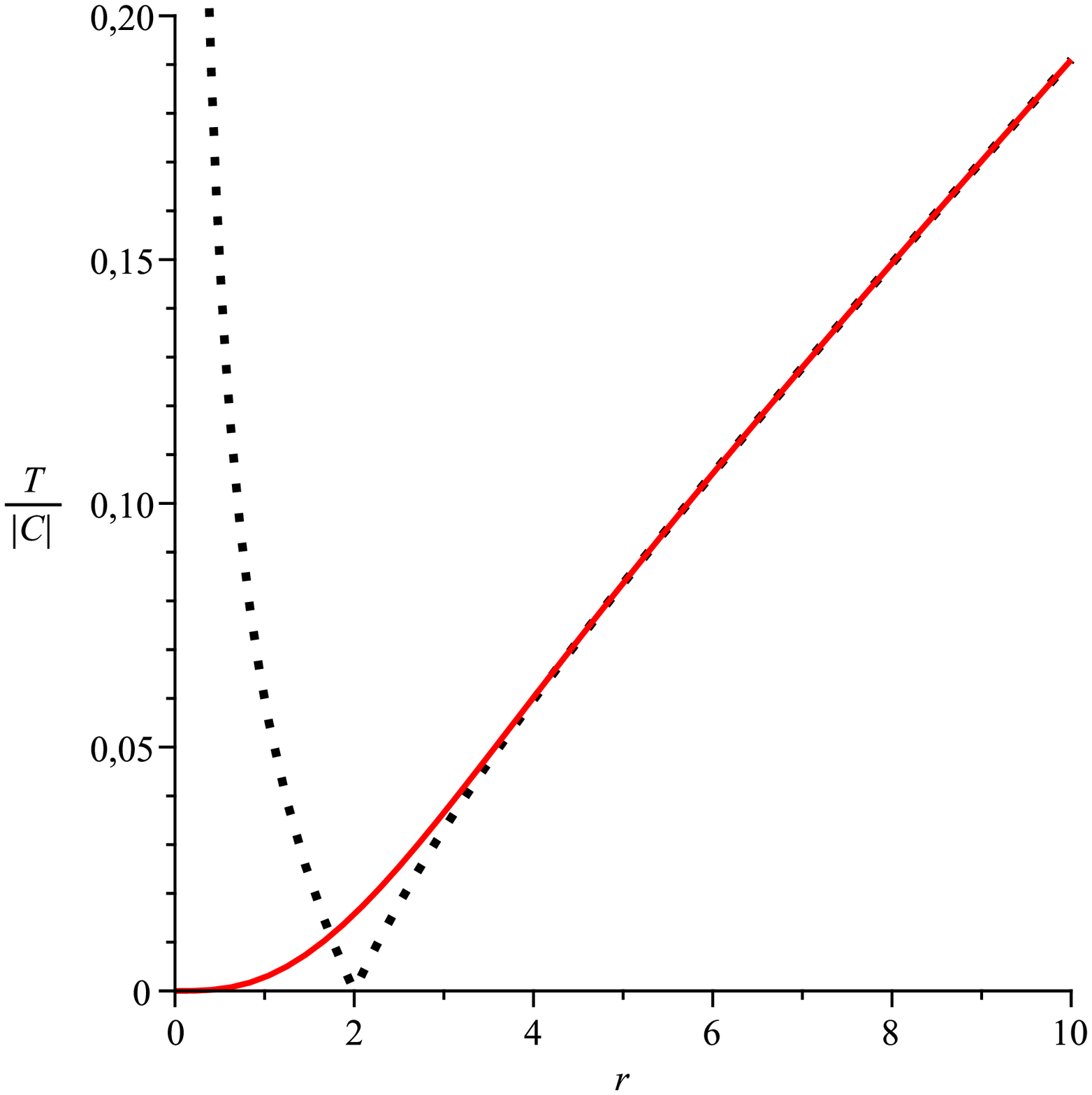}
 \includegraphics[height=3.7cm]{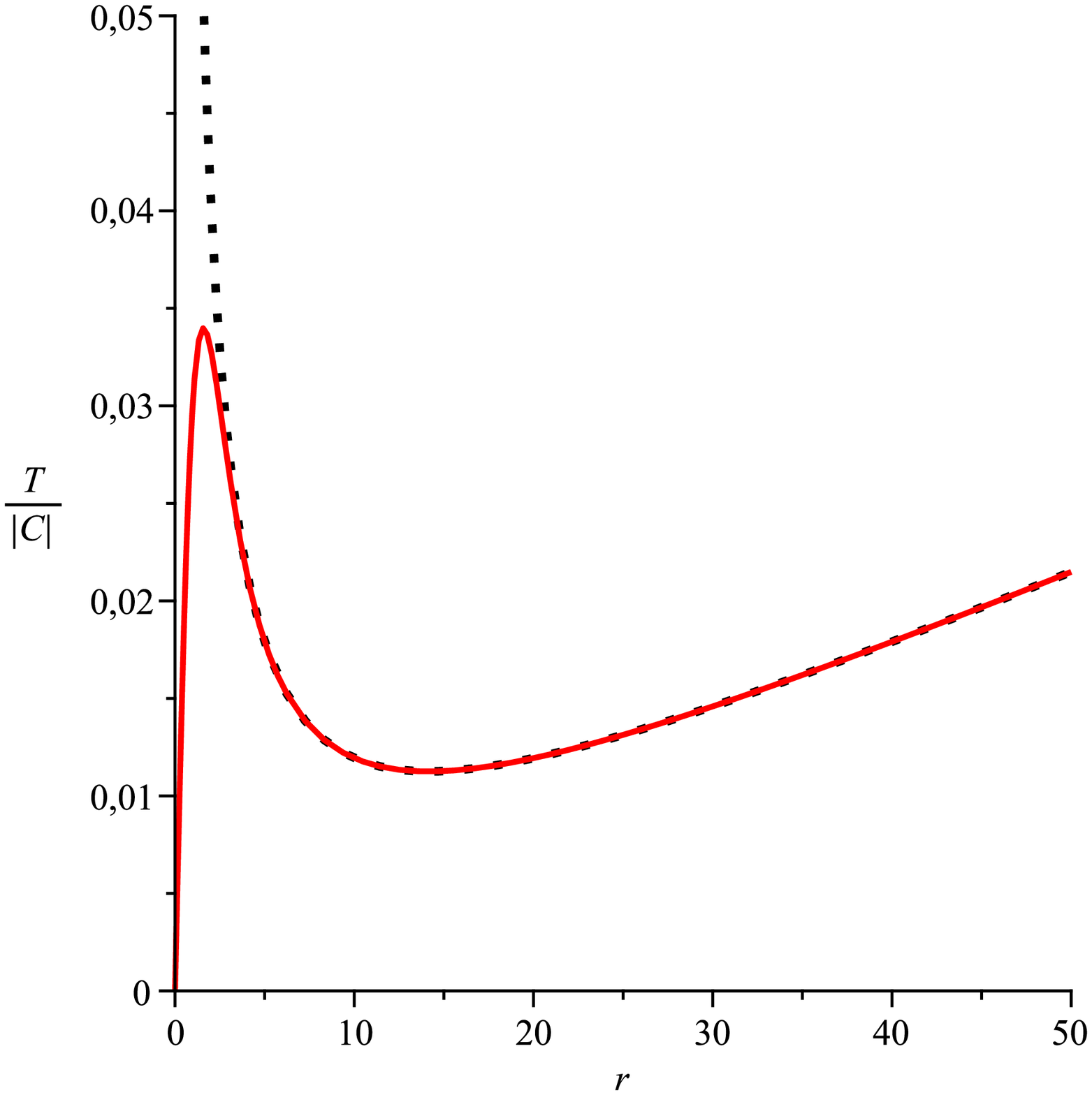}
\includegraphics[height=3.7cm]{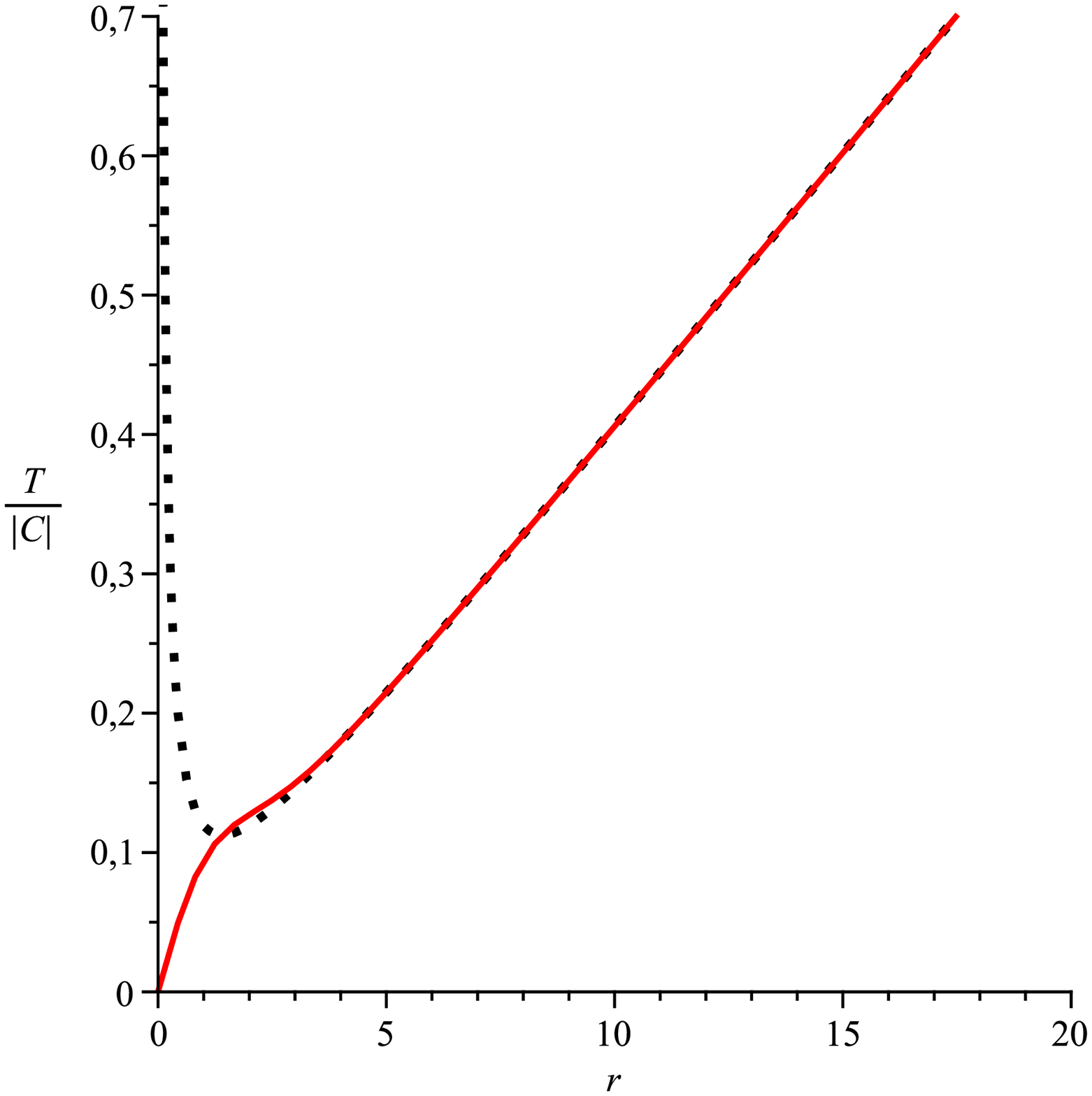}
      \caption{\label{figure10}
Behavior of $T/|C|$ for parameter values (from left):
      $q=5$, $q=\sqrt{2.5}$, $q=\sqrt{2}$ with $\Lambda/C>0$ and $q=10$, $q=1$ with $\Lambda/C<0$.  The dotted plot is the generic commutative solution for the parameters in question.         
      }
     \end{center}
  \end{figure}

The Hawking temperature of the black hole can be readily obtained from Equation~(\ref{ncsolution}).  The periodicity of the complexified temporal component of the metric implies the temperature is

\begin{equation}
T=\frac{\hbar}{2\pi}\left|-\frac{1}{2}\Lambda |x|_>+\frac{1}{2}\left(\frac{1}{2}\Lambda x^2_>+C\right)\frac{f^\prime(|x|_>)}{f(|x|_>)}\right|
\label{nctemp}
\end{equation}
where
\[f(|x|)\equiv 2\sqrt{\frac{\theta}{\pi}}\exp\left(-\frac{x^2}{4\theta}\right)
+\frac{1}{\sqrt{\pi }}  \gamma\left(\half,\frac{x^2}{4\theta}\right)
|x|\]
For $|x|_>\gg \sqrt{\theta}$, {\it i.e.} in the absence of noncommutative effects, the ratio $f^\prime/f=1/|x|_>$ and (\ref{nctemp}) coincides with what found in (\ref{temp}). By an explicit calculation one finds
\[\frac{f^\prime(|x|_>)}{f(|x|_>)}=\frac{1}{|x|_>}\left[1-\frac{\frac{2}{|x|_>}\sqrt{\frac{\theta}{\pi}} \ e^{-x_>^2/4\theta}}{\gamma\left(\half,\frac{x_>^2}{4\theta}\right)+\frac{2}{|x|_>}\sqrt{\frac{\theta}{\pi}} \ e^{-x_>^2/4\theta}}\right]\]
where we have ignored irrelevant signs coming from the derivation of $|x|_>$. It is more convenient to write the above formula as
\beq
\frac{T}{|C|}=\frac{\hbar}{2\pi\sqrt{\theta}}\left|-\frac{1}{2}\left(\frac{\theta\Lambda}{C}\right)r+\frac{1}{2}\left(\frac{1}{2}\left(\frac{\theta\Lambda}{C}\right) r^2+1\right)\frac{f^\prime(r)}{f(r)}\right|
\eeq
with $r=|x|_>/\sqrt{\theta}$. Recalling that when studying horizons we defined $q^2=|C|/\theta|\Lambda|$, we have that for $\Lambda/C>0$ and $q=5$ the scenario is not modified with respect to the usual thermodynamics behavior. Noncommutative corrections occur in a region whose size is smaller than $|x_0|=\sqrt{2 C/\Lambda}$, the size of the black hole remnant, {\it i.e.} the  zero temperature extremal black hole final configuration of the evaporation (see Figure~\ref{figure10}).
As a result the remnant mass has the usual value $M_0=\pm \sqrt{\Lambda C/2}$.
 On the other hand for smaller $q$, the size of the region influenced by noncommutative effects is larger. As a  result the size of the remnant shrinks and the remnant mass decreases too. It is
 \beq
 M_0=\pm\frac{1}{2f(|x|_0)}\left[\half \Lambda (|x|_0)^2+C\right]
 \eeq
For $q\lesssim\sqrt{2}$ the scenario of the final stage of the evaporation changes: in place of the formation of a black hole remnant, the black hole completely evaporate off, giving rise to a regular manifold with a conical singularity at the origin (see Fig. \ref{figure10}).

For $\Lambda/C<0$ we have analogue modifications: instead of a divergent temperature in the terminal phase of the evaporation the hole reaches a zero temperature configuration without leaving any remnant (see Fig. \ref{figure10}).

  \begin{figure}
 \begin{center}
 \includegraphics[height=4.0cm]{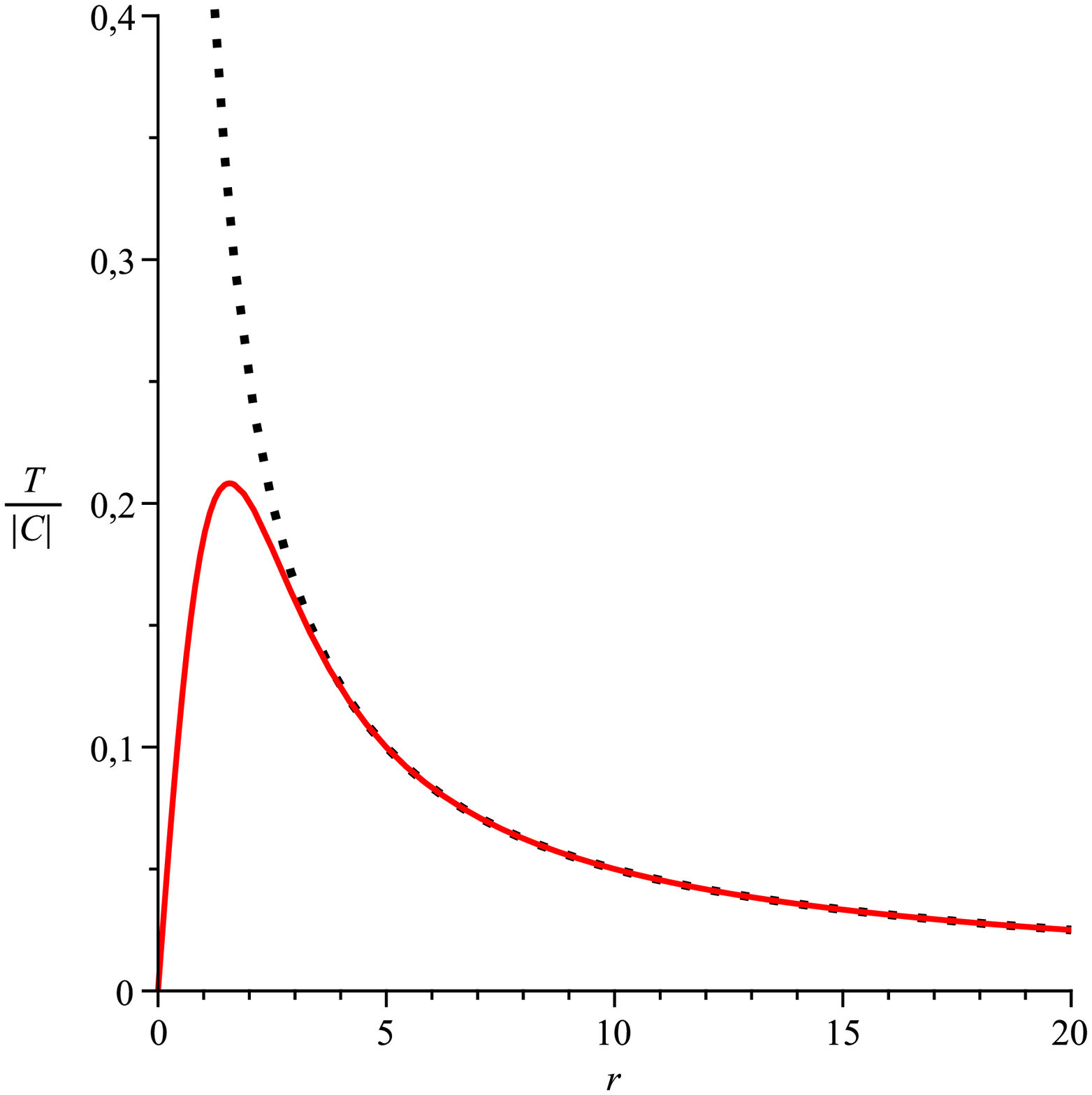}
 \includegraphics[height=4.0cm]{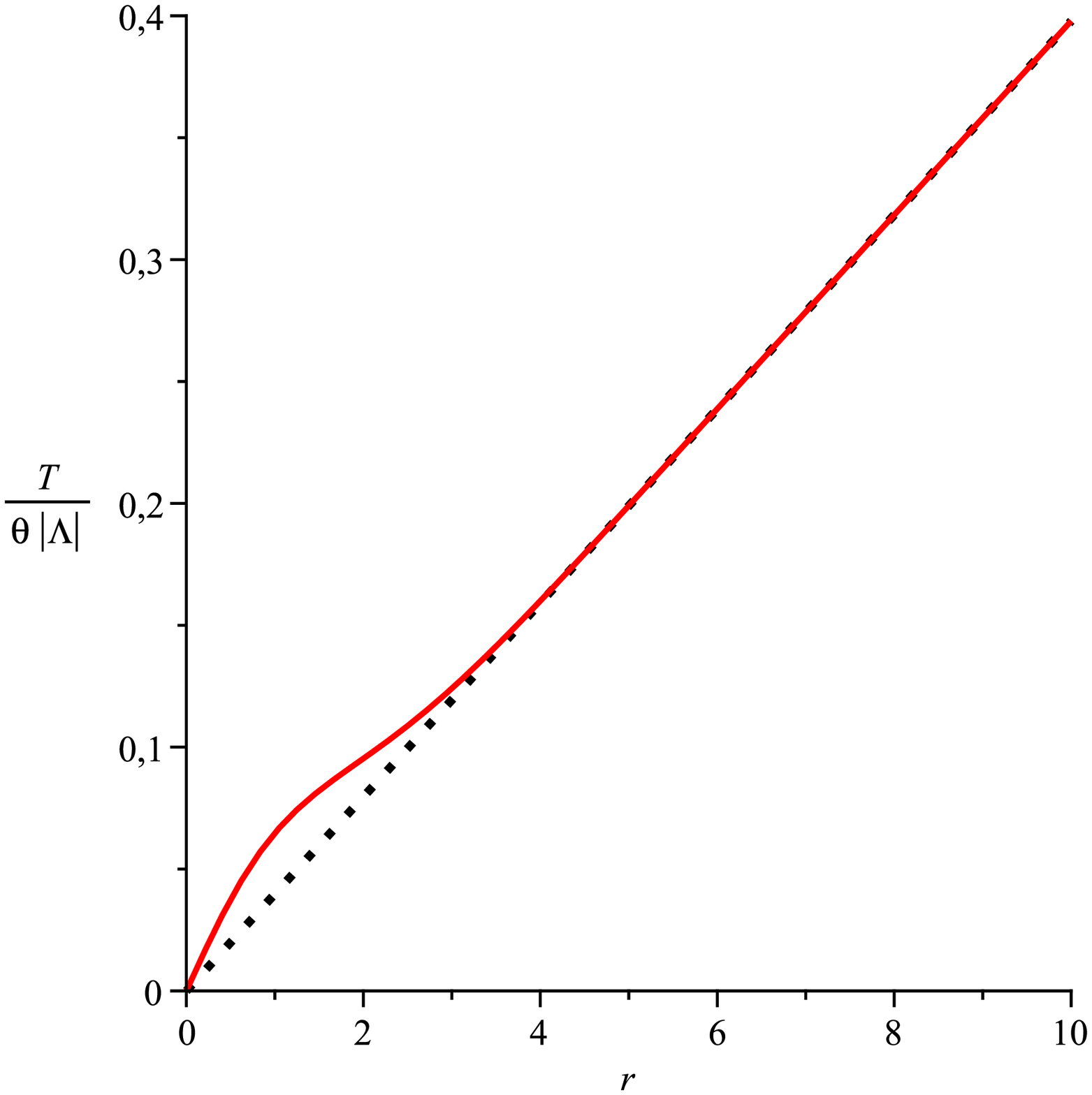}
      \caption{\label{figure11}
Behavior of $T/|\Lambda|\theta$ for parameter values (from left):
      $\Lambda=0$ and $C=0$.  The dotted plot is the generic commutative solution for the parameters in question.          
      }
     \end{center}
  \end{figure}
  
When $\Lambda=0$, we have that the temperature reads
\begin{equation}
T=\frac{\hbar}{2\pi}\left|\frac{1}{2}\ C \ \frac{f^\prime(|x|_>)}{f(|x|_>)}\right|.
\label{nctempL0}
\end{equation}
To plot this function it is useful to write
\beq
\frac{T}{|C|}=\frac{\hbar}{2\pi\sqrt{\theta}}\left|\frac{1}{2}\frac{f^\prime(r)}{f(r)}\right| .
\eeq
This case is particularly interesting since it is the most reminiscent of the four dimensional asymptotically flat   case: as in all the previous noncommutative modified spacetimes, instead of a divergent behaviour of the temperature the black hole undergoes a terminal phase characterized by a positive heat capacity cooling down (see Fig. \ref{figure11}). As already seen in \cite{pn2}, the main difference with respect to four and higher dimensional cases, the black hole completely evaporates without leaving any remnant.  We note the first two solution curves in Fig.~\ref{figure10} are physical only for raddii larger than $r_0$, at which point the evaporation stops.
For smaller values of the horizon radius, the equation $\alpha_\theta(x)=0$ yields no real solutions (and hence no black hole is formed).

Finally, we study the case $C=0$. The temperature is
\begin{equation}
T=\frac{\hbar}{2\pi}\left|-\frac{1}{2}\Lambda |x|_>+\frac{1}{4}\Lambda x^2_> \ \frac{f^\prime(|x|_>)}{f(|x|_>)}\right|
\label{nctempC0}
\end{equation}
which can be conveniently written as
\beq
\frac{T}{|\Lambda|\theta}=\frac{\hbar}{2\pi\sqrt{\theta}}\left|-\frac{1}{2}\ r+\frac{1}{4}\ r^2\ \frac{f^\prime(r)}{f(r)}\right|.
\eeq
From Figure~\ref{figure11}, we see that there are minor modifications with respect the conventional behaviour of the temperature: again the black hole evaporates completely. The only difference is that the final configuration is a regular geometry instead of a singular one.


The entropy of the black hole may be obtained from the differential equation
\beq
 dS =  \frac{dM}{T}.
\eeq
It is more convenient, however, to write the above relation as
\begin{equation}
S=S_0+\int_{|x|_0}^{|x|_>}\frac{1}{T}\frac{\partial M}{\partial |x|_>}d|x|_>
\end{equation}
where $S_0$ is an integration constant and $|x|_0$ is the minimum value of the horizon radius. Depending on the case, {\it i.e.} formation of a remnant or of a conical singularity at the end of the evaporation,  $|x|_0$ can be larger than zero or vanishing.  We start from the internal energy of the black hole, which is nothing but the mass as a function  of the horizon radius
\begin{equation}
M\equiv M(|x|_>)=\frac{1}{2f(|x|_>)} \left(\half \Lambda x^2_>+C\right).
\end{equation}
Thus
\begin{equation}
\frac{\partial M}{\partial |x|_>}=\frac{1}{f(|x|_>)}\left[\half\Lambda |x|_>-\half\left(\half \Lambda x^2_>+C\right)\frac{f^\prime(|x|_>)}{f(|x|_>)}\right].
\end{equation}
As a result the entropy reads
\begin{equation}
S=S_0+\frac{2\pi}{\hbar}\int_{|x|_0}^{|x|_>}d|x|_>\frac{1}{f(|x|_>)}\sgn\left(\half\Lambda |x|_>-\half\left(\half \Lambda x^2_>+C\right)\frac{f^\prime(|x|_>)}{f(|x|_>)}\right).
\end{equation}
In the conventional case, i.e., for $\sqrt{\theta}\to 0$, the function $f(|x|_>)=|x|_>$. As a result one finds
\begin{equation}
S=S_0+\frac{2\pi}{\hbar}\int_{|x|_0}^{|x|_>}d|x|_>\frac{1}{|x|_>}\sgn\left(\half\Lambda |x|_>-
\frac{C}{|x|_>}\right).
\end{equation}
Being $|x|_>\geq|x|_0=\sqrt{2|C/\Lambda|}$, the above formula reduces to
\begin{equation}
S=\frac{2\pi}{\hbar}\ln\left(\frac{|x|_>}{|x|_0}\right)
\end{equation}
where we set $S_0=0$. We notice that also in case of $\sqrt{\theta}\neq 0$, there exists a unique $|x|_0$ such that for $|x|_>>|x|_0$ the temperature is defined, i.e., $T\geq 0$ and the above $\sgn$ function equals $1$ when $\Lambda>0$. In this case $|x|_0$ depends in a more complicated way on $\Lambda$, $C$ and $\theta$. The resulting expression for the entropy formally reads
\begin{equation}
S=\frac{2\pi}{\hbar}\int_{|x|_0}^{|x|_>}d|x|_>\frac{1}{f(|x|_>)}.
\end{equation}
For $\Lambda\leq 0$ and nonnegative $C$ the function $\sgn$ equals $-1$, the entropy can assume negative values.
This is the sign of the presence of \textit{ectropy}, a measure of the tendency of a system to do useful work and reach a more organized configuration. This fact already occurred in the case of $\sqrt{\theta}=0$, i.e., for the conventional metrics \cite{mann2}.
In the absence of black hole remnants, i.e., zeros for the temperature at finite radii, the metric admits a conical singularity at the origin. As a result the entropy reduces to
\begin{equation}
S=\frac{2\pi}{\hbar}\int_{0}^{|x|_>}d|x|_>\frac{1}{f(|x|_>)}\sgn\left(\half\Lambda |x|_>-\half\left(\half \Lambda x^2_>+C\right)\frac{f^\prime(|x|_>)}{f(|x|_>)}\right),
\end{equation}
where we set again $S_0=0$.

\section{Conclusions}

We have shown that the noncommutative effects in $(1+1)-$dimensional black hole physics provide a rich variety of spacetimes with markedly different structure, well beyond the standard case considered in previous works.  With the renewed interest in lower - dimensional gravitation, such behavior is much more than just pedagogical curiosity.   If spacetime really is two-dimensional above some scale that is accessible to current or near-future experiments ({\it e.g.} 100~TeV or more as suggested \cite{jrmds}), the unique thermodynamics discussed herein can lead to an equally novel array of testable phenomenology.    Higher-energy collisions could lead to tell-tale quantum black hole signatures that would distinctly identify the model, much in the same way as discussed in \cite{landsberg}. 
The model-dependence of (1+1)-dimensional noncommutative black hole phenomenology
({\it e.g.} in the CGHS framework) is also of possible future interest.
 The implications for early-universe physics is profound: promordial black hole remnants in a two-dimensional early universe can significantly affect the initial conditions for matter distribution and dark matter densities, an effect which could in principle be ``etched'' in the cosmic microwave background. 

\vskip 2cm
\noindent{\bf Acknowledgments}\\
JRM and PN would like to thank the Perimeter Institute for Theoretical Physics, Waterloo, ON, Canada for the kind hospitality during the initial period of work on this project. This work is supported by the Helmholtz International Center for FAIR within the
framework of the LOEWE program (Landesoffensive zur Entwicklung Wissenschaftlich-\"{O}konomischer Exzellenz) launched by the State of Hesse and in part by the European Cooperation in Science and Technology (COST) action MP0905 ``Black Holes in a Violent Universe''.
The authors would like to thank M. Rinaldi for valuable comments to the manuscript.

 \end{document}